\documentclass[%
 aip,
amsmath,
amssymb,
nofootinbib,
reprint,%
floatfix,
]{revtex4-1}
\usepackage{subcaption}
\usepackage{graphicx}% Include figure files
\usepackage{dcolumn}% Align table columns on decimal point
\usepackage{bm}% bold math
\usepackage[utf8]{inputenc}
\usepackage[T1]{fontenc}
\usepackage{mathptmx}
\usepackage{etoolbox}
\usepackage{comment}
\usepackage{booktabs}
\usepackage{tabularx}
\newcolumntype{Y}{>{\centering\arraybackslash}X}

\usepackage{tikz}
\usepackage{pgfplots}
 \pgfplotsset{
        layers/my layer set/.define layer set={
            background,
            main,
            foreground
        }{
        },
        set layers=my layer set
    }
\usepackage{float}
\makeatletter
\let\newfloat\newfloat@ltx
\makeatother
\usepackage{algorithm}
\usepackage{algorithmicx}
\usepackage{algpseudocode}

\DeclareMathOperator*{\argmin}{arg\,min}
\makeatletter
\def\@email#1#2{%
 \endgroup
 \patchcmd{\titleblock@produce}
  {\frontmatter@RRAPformat}
  {\frontmatter@RRAPformat{\produce@RRAP{*#1\href{mailto:#2}{#2}}}\frontmatter@RRAPformat}
  {}{}
}%
\makeatother
\begin{document}

\preprint{AIP/123-QED}

\title[Experimental jet control with Bayesian optimization and persistent data topology]{Experimental jet control with Bayesian optimization and persistent\\ data topology}%{Experimental mixing enhancement by radial acoustic excitation for \\ a turbulent jet with machine learning}
% Force line breaks with \\
\author{Johann Moritz Reumsch\"ussel}%
\homepage{These authors contributed equally to the work.}
\affiliation{Chair of Fluid Dynamics, Technische Universität Berlin, Germany
}
\author{Yiqing Li}
\homepage{These authors contributed equally to the work.}
\affiliation{ Chair of Artificial Intelligence and Aerodynamics, 
School of Mechanical Engineering and Automation, Harbin Institute of Technology, 518055 Shenzhen, P.~R.~China
}
\author{Philipp Maximilian zur Nedden}
\affiliation{Chair of Fluid Dynamics, Technische Universität Berlin, Germany
}
\author{Tianyu Wang} 
\affiliation{ Chair of Artificial Intelligence and Aerodynamics, 
School of Mechanical Engineering and Automation, Harbin Institute of Technology, 518055 Shenzhen, P.~R.~China
}
\author{\\Bernd R. Noack}
\email{bernd.noack@hit.edu.cn, oliver.paschereit@tu-berlin.de}
\affiliation{ Chair of Artificial Intelligence and Aerodynamics, 
School of Mechanical Engineering and Automation, Harbin Institute of Technology, 518055 Shenzhen, P.~R.~China
}
\affiliation{Guangdong Provincial Key Laboratory of Intelligent Morphing Mechanisms and Adaptive Robotics,
Harbin Institute of Technology, 518055 Shenzhen, P.~R.~China
}
\affiliation{Chair of Fluid Dynamics, Technische Universität Berlin, Germany
}
\author{Christian Oliver Paschereit}
\affiliation{Chair of Fluid Dynamics, Technische Universität Berlin, Germany
}%\email{bernd.noack@hit.edu.cn}

\date{\today}% It is always \today, today,

\begin{abstract}
This study experimentally optimizes the mixing of a turbulent jet at Reynolds number $10000$ with the surrounding air by targeted shear layer actuation.
The forcing is composed of superposed harmonic signals of different azimuthal wavenumber $m$ generated by eight loudspeakers circumferentially distributed around the nozzle lip. Amplitudes and frequencies of the individual harmonic contributions serve as optimization parameters and the time-averaged centerline velocity downstream of the potential core is used as a metric for mixing optimization.
% --- Problem ---
The actuation is optimized through Bayesian optimization. Three search spaces are explored --- axisymmetric forcing, $m=0$, superposed axisymmetric and helical forcing, $m \in \{0,1\}$, and axisymmetric actuation combined with two counter-rotating helical modes, $m \in \{-1,0,1\}$. High-speed PIV is employed to analyze the jet response to the optimized forcing. The optimization processes are analyzed by persistent data topology.
In the search space of axisymmetric excitation, the routine identifies an actuation at the natural frequency of the flow to be most efficient, with the centerline velocity being decreased by $15\%$.
The optimal solutions in both the two-mode and three-mode search space converge to a similar forcing with one axial and one helical mode combined at a frequency ratio of around $2.3$. Spectral analysis of the PIV images reveals that for the identified optimal forcing frequencies, a non-linear interaction between forced and natural structures in the jet flow is triggered, leading to a reduction in centerline velocity of around $35\%$.
The topology of the most complex search space from the discrete data reveals four basins of attractions, classified into three forcing patterns including axisymmetric, axisymmetric-helical, and axisymmetric-flapping.
Two deep basins are related to the optimal pattern found as axisymmetric-helical, and the others are shallower.
\end{abstract}

\maketitle

\section{Introduction}
% Intro
The axisymmetric turbulent jet represents a fundamental configuration of shear flow and has attracted significant attention in fluid dynamics research. Moreover, its practical significance extends to various technical applications, such as gas turbine combustion~\cite{beuth2024thermoacoustic} or chemical process reactors~\cite{santos2013state}. When a directed jet flow emerges into the surrounding fluid, a core persists along the central axis in which the axial velocity is initially maintained. The core region is surrounded by a shear layer where the velocity gradient causes significant interaction between the jet and the surrounding fluid, resulting in the formation of unsteady phenomena~\cite{zaman1980jfm, Cohen1987a}. Vortices are formed in the shear layers due to Kelvin-Helmholtz type instabilities. This process causes the surrounding fluid to mix and entrain with the jet, resulting in an increase of jet flux in the direction of flow. Ordered structures in the turbulent motion of jets were studied by \citet{crow1971jfm}, who discovered periodic fluctuation at specific Strouhal numbers in the shear layers. The properties of these coherent structures, both in natural flow and in externally excited form, have been investigated in many respects since then. The high sensitivity of the shear layer in the region of strong velocity gradients in the vicinity of the nozzle lip has been found to lead to very efficient control measures\cite{Michalke1965}. 

% Jet control
Various types of actuators were used in this region to control the flow.
For example, dual loudspeakers were employed for transverse actuation in the study of \citet{parekh1989phd} alongside a single upstream one. \citet{wiltse1993manipulation} placed four resonantly driven piezoelectric actuators along the sides of a square exit. \citet{parekh1996innovative} extended the use of piezoelectric actuators to supersonic jet flows with piezoelectric wedge actuators. They also studied pulsed and synthetic excitation by proposing two slot-jet actuators on opposite sides.
\citet{juvet1994phd} designed a Coanda effect-controlled jet, in which tangential blowing at the throat is used to support the boundary layer along a turn.
\citet{suzuki1999symposium} mounted miniature electromagnetic flap actuators on the periphery of a coaxial nozzle.
\citet{utkin2006jfd} used eight arc filament plasma actuators for jet control. \citet{wu2018ef} utilized six independently controlled unsteady minijets and \citet{segawa2008wall} investigated the jet control using a Doughnut-shaped DBD actuator.

% forced response research
For harmonic monofrequent axisymmetric excitation, forcing frequencies close to the dominant natural frequency of the unforced flow have proven to be particularly efficient~\cite{crow1971jfm, boguslawski2019jfm}. Non-axisymmetric excitation patterns were tested by means of several circumferentially distributed actuators, which are controlled at the same frequency but with a specified time delay. Due to the propagation of the excited structures, helical vortex ropes can be generated. Interaction between helical structures were studied by \citet{long_controlled_1992}. A significant distortion of the cross-sectional shape of the mean jet flow by the combination of excitation patterns of different azimuthal order was shown experimentally by \citet{paschereit_experimental_1995}. A particular jet structure, occurring as a result of combined actuation of helical and axial modes known as blooming jets, was found by \citet{reynolds_bifurcation_2003}.
% data-driven approach for flow control
In most of these studies, theoretical models are compared with experimental observations in a classical approach; however, with today's computing power, data-driven approaches can be used for the investigation of flow control~\cite{Brunton2020}. This involves using efficient mathematical routines to iteratively adjust the excitation parameters in order to optimize a target metric. 
% data-driven methods applied to jet flow control
Optimization of control in jet flows has been conducted in both numerical and experimental studies.
Using vortex models, \citet{Koumoutsakos2001aiaaj} studied the best combination of three actuation modes imposed in the form of radial excitation at the jet outlet. Only one wave dominates in the final solution. Combining an axial and a helical mode in direct numerical simulation,
\citet{hilgers2001fdr} performed an optimization of the Strouhal numbers to maximize the scalar dissipation of the produced bifurcating jet. The optimized frequencies correspond approximately to the natural Strouhal number of the jet and twice its value. \citet{zhou2020jfm} optimized closed-loop control laws of unsteady minijet actuators for mixing improvement. Zur Nedden \textit{et al.}\cite{zurNedden_AIAA} built data-driven surrogate models to study the effect different forcing patterns for jet control. % with Genetic programming
\citet{zigunov2022jfm} performed a search of the optimal actuator locations and parameters of mini-jets for active jet noise reduction. %with Genetic algorithm 
Different efficient algorithms allow even highly complex flow control configurations to be optimized~\cite{li2022jfm, blanchard2021ams, pino2023jfm}. For jet control, evolutionary strategies have been employed in most of the studies~\cite{Koumoutsakos2001aiaaj,zhou2020jfm,zigunov2022jfm}, which are known to converge robustly to global maxima, yet at the cost of many required function calls~\citep{eiben2003introduction}. A methodology that is known to identify optima based on only a few function calls is Bayesian optimization (BO)~\cite{mockus1975bayesian}. BO is different from evolutionary strategies that iterate by population. In each step of Bayesian optimization, a surrogate model, mostly in the form of a Gaussian Process (GP) is fitted to the available data, which allows probabilistic target predictions in the entire parameter space~\cite{Rasmussen2006}. The model predictions are then used to guide the selection of subsequent parameter combinations for testing. For this purpose, an acquisition function serves as a sampling criterion, aiming to strike a balance between exploration of areas of the search space where the function is poorly understood and exploitation of the surrogate to sample in areas where optima can be expected based on available knowledge. Various acquisition functions have proven highly effective to fulfill this purpose like the Expected Improvement\citep{Jones1998} or the Lower Confident Bound acquisition function\citep{srinivas2009gaussian}, among many others\citep{FORRESTER200950, hennig2012entropy, blanchard2021jcp}. BO is advantageous for optimizing functions that are costly and therefore limited in the number of calls~\cite{shahriari2015taking}. Accordingly, it has been found suitable for a wide range of fluid dynamics problems like meteorological applications~\cite{garnett2010bayesian}, the layout of wind farms\cite{Bempedelis2023} and combustor design for gas turbines\cite{reumschuessel2024}, among many others~\cite{MORITA2022110788}.

In this study, we investigate to what extent Bayesian optimization can be used to optimize the excitation of a jet flow with the goal of generating physical insight. We combine concepts from classical jet research with machine learning by applying BO to the parameters of superposed actuation patterns which have been extensively studied in the classical literature. We investigate three design spaces which combine axisymmetric and helical forcing patterns. Therefore we make use of a jet nozzle equipped with an array of azimuthally distributed loudspeakers to acoustically force the flow. The design is similar to the ones employed to study the interaction between two counter-rotating helical waves in the free jet by \citet{long1992jfm},
and in different studies on stability analysis of swirling jets \cite{panda1994pf,oberleithner2012phd, Mueller_2020}. We link the optimal excitations found by BO with the natural structures occurring in the flow and analyze their mechanism of action. Furthermore, we use all points queried during the optimizations to perform a topological analysis of the most complex search space. With the help of multidimensional scaling, we create two-dimensional proximity maps and identify basins of attraction, which also allows us to identify suboptimal local minima in the forcing space.

This paper is organized as following:
The optimization problem is introduced in \S~\ref{ToC:setup}, including a description of the experimental setup and the flow measurement, the definition of the excitation space, and the objective function. The optimization strategy is detailed in \S~\ref{ToC:bo_and_PDT}. This part also includes details about the multidimensional scaling methods used for topological data analysis. 
Finally, we discuss the learning process, the optimized control schemes, and the resulting flow fields in \S~\ref{ToC:res}. We also perform a topological analysis of the design space and analyze the attraction properties of the previously found minima. The findings are summarized in \S~\ref{ToC:concl}.
\vspace{-5mm}

\section{Problem setup}
\label{ToC:setup}
\subsection{Experimental setup}
% --------------------- start figure ---------------------
\begin{figure}
    \centering
    \includegraphics[width=1\linewidth]{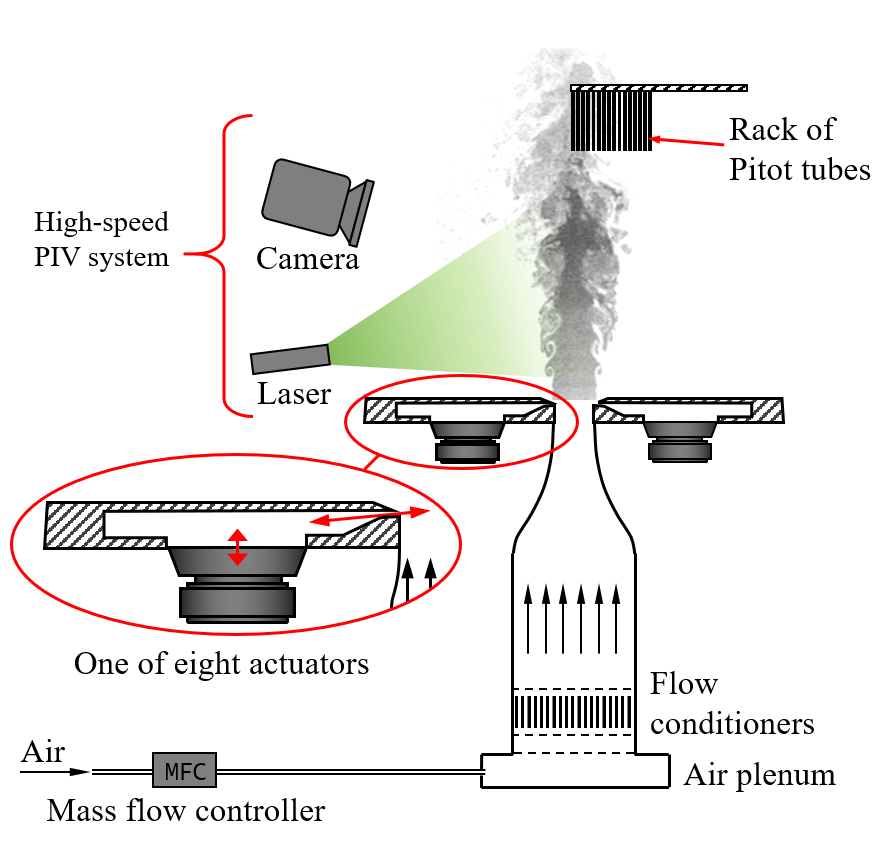}
    \caption{Sketch of the jet plant with details on instrumentation and forcing devices.}
    \label{fig:jet}
\end{figure}
% --------------------- end figure ---------------------
The flow configuration under investigation is a round, non-swirling air jet at Reynolds number $10^4$ based on the nozzle diameter $D_j=51$mm and jet velocity $u_j= 2.93$m/s, where $u_j$ is calculated from the mass flow $m_j$, nozzle cross sectional area $A_j= \frac{\pi}{4}~D_j^2$ and density $\rho$. The experimental setup is sketched in Fig.~\ref{fig:jet}. A thermal mass flow controller is used to control the air flow, which is fed to a plenum and passed through flow straighteners before reaching the contracting nozzle and finally emerging into the quiescent ambient air. In the shear layers between the jet and the surrounding, ring-shaped vortices emerge as a result of Kelvin-Helmholtz-type instabilities. These cause the jet to entrain the surrounding fluid, so that the mass flow in the direction of flow increases. 

To monitor the flow for optimization, we use a rake of $16$ Pitot tubes which are oriented in flow direction and are located $6.5D_j$ downstream of the nozzle at different radial locations from where they transfer the pressure to piezoelectric differential pressure sensors that capture the stagnation pressure $p_{s}$. The other side of the sensors are connected to a still plenum and measure the static pressure $p_{stat}$, so that axial velocity profiles can be calculated from the measured pressure differences,
\begin{equation}
    u_x = \sqrt{2(p_{s} - p_{stat})/\rho}.
\end{equation}
Additional measurements of the flow were performed after the optimization process by means of constant temperature anemometry (CTA). The utilized hot wire probe features a tungsten wire of 5$\mu$m (Dantec), and is connected to an IFA 100 CTA bridge, which regulates the supply voltage to the probe in order to keep the electrical resistance constant. Using King's law with parameters calibrated prior to measurements the supply voltage is converted to flow velocity. The probe is attached to a two-axis traversing system and is used to capture profiles of time-averages of the velocity and of fluctuations in form of root mean square values. The hot wire probe is also employed to calibrate the forcing device, as described below.
To gain further insight into coherent structures in the flow, high-speed Particle Image Velocimetry (PIV) is used to record time-resolved flow fields. For this purpose, silicone oil droplets are introduced into the flow downstream of the mass flow controller. A high-speed camera (Photron SA-Z) records pairs of images of the droplets in the flow illuminated by a nd:ylf laser (Quantronix Darwin-Duo) in a section plane across the jet. The double images are processed into snapshots of the flow field using standard PIV processing~\cite{willert1991digital} by means of the software package PIVview. Images are recorded with 30$\mu$s acquisition time and the PIV system is operated at a sampling frequency of 1kHz.

\subsection{Flow excitation} \label{Problem:inputs}
In the vicinity of the nozzle exit, where the shear layer is highly susceptible to perturbations, the flow is excited in order to optimize the mixing between jet and surrounding fluid. Therefore, acoustic actuation is applied at the nozzle lip using an array of eight loudspeakers equally spaced in an azimuthal arrangement around the nozzle. An acoustic wave-guide from each actuator terminates in a rectangular duct leading to a narrow slit of height $h_a=1$mm, where operation of the speakers generates oscillating zero-net-mass-flow jets, as detailed in Fig.~\ref{fig:jet}. The forcing slits are designed such that they do not interfere with the main jet flow when the speakers are inactive. In order to exclude the forcing device's characteristic response from the optimization process, the actuators are calibrated frequency-dependent to the velocity oscillations generated at the slit-outlet $\hat{u}_{ac}$, measured using the hot-wire probe. Therefore, the actuation mass flux $m_{ac}$ is constrained to $1\%$ of the jet mass flux for each forcing frequency, such that
\begin{equation}
\label{eq:massflowratio}
    \frac{m_{ac}}{m_j} = \frac{A_{ac} \hat{u}_{ac} }{A_{j} u_j} \le 1 \% \quad  \forall \omega, 
\end{equation}
where $m_{ac}$ is calculated from the slit velocity amplitude $\hat{u}_{ac}$ and $A_j$ and $A_{ac} = \pi D_j h_a$ are the jet nozzle and forcing outlet cross sectional areas, respectively. The constraint in Eq.~\eqref{eq:massflowratio} yields a maximum for the actuation velocity contribution at every frequency
\begin{equation}\label{eq:maxvel}
    \hat{u}_{ac} \leq u_{1\!\%} = 0.9221~\text{m/s} \quad \forall \omega, 
\end{equation}
from which the frequency-dependent maximum supply voltage for the speakers was determined.

The loudspeakers are connected to individual amplifiers which are driven by a set of digital–to–analog converters under program control. The speakers are additionally adjusted to equal amplitudes under no–flow conditions using a microphone located at the centerline in the exit plane of the nozzle. 

The actuation signals are composed of superposition of monofrequent harmonic signals. By varying the phase difference between the actuators, forcing patterns of different azimuthal orders can be generated, the amplitudes and frequencies of which serve as optimization parameters. The actuation velocity signal generated by speaker $i$ is the superposition of forcing modes at order $m$ and reads:
\begin{equation}
    u_{ac,i}(t) = \sum_{m=-4}^{4} \underbrace{a_m~u_{1\!\%}}_{\hat{u}_{ac}}~\sin\Big(\underbrace{St_m \omega_c}_{\omega_m} t + \frac{i m \pi}{4}\Big),
\end{equation}
where $\omega_m$ is the angular frequency, which we express in terms of Strouhal numbers $St_m = \omega_m D_j / 2 \pi u_j$ and characteristic angular frequency $\omega_c = 2\pi u_j/D_j$ and the associated velocity amplitude is expressed in terms of the constraint value from Eq.~\eqref{eq:maxvel} and a non-dimensional factor $a_m$, ranging from 0 to 1. By considering different superposed patterns, optimization problems of varying complexity are generated. With an array of eight actuators the highest azimuthal mode numbers that can be excited are $m = \pm 4$. However, within the scope of this study the azimuthal order was limited to $m\in\{-1,0,1\}$. This leaves the actuation with up to $6$ degrees of freedom $\bm b = [b_1,..,b_6]$, including $3$ amplitudes $a_m$ and  $3$ frequencies $\omega_m$,
\begin{equation}\label{eq:parameters}
    \bm{b} = [a_{m}, \dots, \widetilde{St}_m, \dots], \quad m\in\{-1, 0, 1 \}.
\end{equation}
Note that within the optimization, normalized parameters $a$ and $\widetilde{St}$ are considered which range between 0 and 1, according to the constraints. In regard of the amplitudes, the parameter space $\bm{B}$ is restricted to a maximum actuation mass flux of 1\%, as defined in Eq.~\eqref{eq:massflowratio}. The considered frequencies are constrained to $0.1< St < 1$ such that $St_m = 0.1+(1-0.1)\widetilde{St}_m$. 

\subsection{Objective function}
\label{Problem:cost}
We intend to find optimal forcing schemes that maximize the generation of vortices and thus  improves mixing between the jet and the ambient fluid. This leads to an increased growth of shear layers in flow direction, stronger jet expansion and a shorter potential core. To assess different forcing configurations in this regard, the velocity data from the rack of Pitot tubes is evaluated. 
The forcing is optimized by minimizing the time-averaged axial velocity $\bar{u}_{cl}$ at a fixed position on the centerline which is a measure for the shortening of the potential core, as has been done in different other studies~\citep{nathan2006prog, wu2018ef}. The measurement position $x_0$ is chosen to be $6.5D_j$ downstream of the jet exit, and thus slightly behind the potential core which ends around $x_p=6D_j$ in the natural flow.
To avoid errors in the optimization due to drifts in fluid properties over time, the optimization quantity is normalized by the value without forcing $\bar{u}_{cl,0}$, which is re-measured every 20 iterations,
\begin{equation}\label{eq:J}
     J(\bm b)=\bar{u}_{cl,x_0}(\bm b)/\bar{u}_{cl,0}.
\end{equation}
We also tested other informative objective metrics that represent a larger jet spread and incorporate sensor information from all 16 Pitot tubes to increase robustness. However, we found the centerline to be superior in terms of reproducibility and required measurement time. A measurement time of 20~seconds was found to be sufficient for a deviation below 1\% between repeated experiments. More details on both other target metrics tested and the reproducibility study are given in the appendix.
\section{Methodology}  \label{ToC:bo_and_PDT}
 The task of mixing enhancement is formulated as a minimization problem of the objective function $J\in\mathbb{R}$ with respect to the parameter input vector 
 $\bm{b}\in \bm{B} \subset \mathbb{R}^d$:
\begin{equation} \label{eq:opti_problem}
    \bm{b}^* = \underset{\bm{b}\in{\bm{B}}}{\argmin}~J(\bm{b}).
\end{equation}
As discussed in Sec.~\ref{Problem:cost}, the objective function is related to the streamwise velocity at the centerline, $6.5D_j$ downstream of the nozzle exit, Eq.~\eqref{eq:J}, and the actuation vector spanning the design space $\bm{B}$ contains the amplitudes and frequencies of forcing modes of different azimuthal order, Eq.~\eqref{eq:parameters}. We explore three different design spaces by means of data-driven optimization. The design spaces are generated by superposition of actuation at different azimuthal mode number. First, we consider only axisymmetric forcing, $\bm{b}\in \bm{B}_1 \subset \mathbb{R}^2$, $m=0$. Subsequently, we extend the forcing to also enable helical actuation $\bm{b}\in \bm{B}_2 \subset \mathbb{R}^4$, $m\in\{0, 1 \}$. Finally, we investigate a design space that allows axial, helical and counter-rotating helical excitation, $\bm{b}\in \bm{B}_3 \subset \mathbb{R}^6$, $m\in\{-1, 0, 1 \}$. Thereby we increase the dimension of the parameter space gradually from $d=2$ to 6.

After having identified optimal forcing patterns in all three design spaces by applying Bayesian optimization, we augment the queried data from $\bm{B}_1$ and $\bm{B}_2$ to the dimension of $\bm{B}_3$, such that a comprehensive data set is created from all queried points. This allows us to perform a topological analysis of the most extensive design space $\bm{B}_3$ by means of classical multi-dimensional scaling.

\subsection{Bayesian optimization}
 \label{ToC:bo}
% -------------------- Begin algorithm --------------------
\begin{algorithm}[b]
    \caption{Bayesian optimization of jet control}
    \label{alg:bo}
    % \algsetup{linenosize=\tiny}
    \small
    \begin{algorithmic}[1]
        % \Require $n \geq 0$
        % \Ensure $y = x^n$
        \State Input: maximum iterations $N_{max}$.
        \State Initialize: dataset $\mathcal{D} =\{\bm{b}_i, J_i\}_{i=1}^{n_{0}}$, iteration $n=0$.
        \While{$n \leq N_{max}$}
            \State Sample actuation: $\bm{b}_n = \argmin_{\bm{b} \in \bm{B}} \alpha(\bm{b}, \bar{f}, \mathcal{D})$
            \State Perform experiment: $J_n = J(\bm{b}_n)$
            \State $\mathcal{D} \gets \mathcal{D} \cup \{\bm{b}_n, J_n\}$
            \State Update surrogate: $\bar{f}$
            \State $n \gets n+1$
        \EndWhile
            \State Output: optimized jet flow actuation $\bm{b}^\star$.
    \end{algorithmic}
\end{algorithm}   
% -------------------- End algorithm --------------------
The optimization problem Eq.~\ref{eq:opti_problem} is tackled using Bayesian optimization. The procedure of the adopted approach is shown in Alg.~\ref{alg:bo}.
To initialize the process, a starting dataset $\mathcal{D}$
% denoted as $\mathcal{D}_0$, 
comprising of $n_0 = d+1$ input-output pairs $\{\bm{b}_i, J_i \}_{i=1}^{n_0}$ is generated by randomly sampling from $\bm{B}$ and performing $n_0$ experiments.
From the dataset, a surrogate model $\bar f$ is trained to approximate the latent objective function $J$. 
The dataset is then enriched with a new data point $\{\bm{b}_n, J_n\}$, the choice of which is directed by an acquisition function $\alpha(\bm{b}, \bar f, \mathcal{D})$. New points are initially selected to enhance the accuracy of the surrogate model across the search space $\bm{B}$ which is later exploited to sample around predicted minima of $J$.
As soon as the query budget is met, $n=N_{max}$, 
the algorithm terminates and returns the best design vector $\bm{b}^\star$ tested during the process.

The surrogate model $\bar{f}(\bm{b})$ is built by a Gaussian process and follows a normal distribution with posterior mean $\mu(\bm{b})$ and variance $\sigma^2(\bm{b})$, conditioned on available data $\mathcal{D}=\{\mathbf{B}, \mathbf{q}\}$~\cite{Rasmussen2006}:
\begin{equation}
    \mu(\bm{b}) = m_0 + k(\bm{b}, \mathbf{B})K^{-1}(\mathbf{q} - m_0),
\end{equation}
\begin{equation}
    \sigma^2(\bm{b}) = k(\bm{b}, \bm{b})-k(\bm{b}, \mathbf{B})\mathbf{K}^{-1}k(\mathbf{B}, \bm{b}).
\end{equation}
Here $m_0$ is the constant mean, and $k(\bm{b}, \bm{b}^\prime;\bm{\Theta})$ is the covariance with $\bm{\Theta}$ as hyper-parameters. $m_0$ and $\bm{\Theta}$ are calibrated to the available data using maximum-likelihood estimation\cite{Rasmussen2006}. $\mathbf{B} \in \mathbb{R}^{d \times n}$ contains the tested actuation vectors, and $\mathbf{q} \in \mathbb{R}^n$ the corresponding objective values.
Using the available data set $\mathcal{D}$, the prediction is given by the surrogate model's posterior mean $\mu$ with variance $\sigma^2$ as uncertainty score. To guide the selection of data for the subsequent iteration within the control space the likelihood-weighted lower confidence bound acquisition function \cite{blanchard2021jcp} is employed:
\begin{equation}
    \label{equ:acquisition}
    \alpha(\bm{b}) = \mu(\bm{b}) - \kappa \sigma(\bm{b}) w(\bm{b}), ~ w(\bm{b}) = \frac{p_{\bm{b}}(\bm{b})}{p_\mu(\mu(\bm{b}))}.
\end{equation} 
Here, $\kappa$ is chosen as $1$ to balance exploration and exploitation.
The likelihood ratio $w(\bm{b})$ weights the input density $p_{\bm{b}}$ and the output density $p_\mu$ to measure the uncertainty of point $\bm{b}$ against its expected impact on the cost function.

\subsection{Persistent data topology}
\label{ToC:PDT}
Persistent data topology (PDT) is used as a data analysis tool to identify cost function minima and their persistence to noise \cite{wang2023book, WangTY2023}. 
In the context of flow control optimization, PDT was recently employed in a numerical study by \citet{li2023jfm} to extract effective mixing flow patterns, like blooming, flapping, and helical structures in 22-dimensional forcing spaces. Using a similar methodology, the current study uses PDT for the identification of local minima within the discrete data queried during optimization.
%Within available data points from the optimization dataset $\mathcal{D}\subset \mathbb{R}^{d}$ , $\bm{b}^\circ$ is identified as a local minimum if
%\begin{equation}
%    J(\bm{b}^\circ) \leq J(\bm{b}) \;  \forall \;  \bm{b}\in\mathcal{D}_0, 
%\end{equation}
%where the subset $\mathcal{D}_0 \subset \mathcal{D}$ is a neighborhood of $\bm{b}^\circ$.
%The neighbourhood is set to be composed of the $K$ nearest neighbors of $\bm{b}^\circ$, $\bm{b}^{k_1}, \dots, \bm{b}^{k_K}$, which are identified from sorting all available data points according to their distance to $\bm{b}^\circ$, 
%\begin{equation}
%   \| \bm{b}^{k_1}-\bm{b}^\circ \| \leq \| \bm{b}^{k_2}-\bm{b}^\circ \| \leq \cdots \leq \| \bm{b}^{k_{M-1}}-\bm{b}^\circ \|.
%\end{equation}
%Here $\|.\|$ denotes the Euclidean distance in $d$ dimensions and $K$ is chosen to be $d+1$.
%\begin{comment}
%, $\{k_1, k_2, \cdots, k_{M-1}\} \in \{1, 2, \cdots, M-1\}$.
%Here, $K$ is chosen to be $N+1$, where $N$ is the dimension of $\bm{b}$.
%For the point $\bm{b}$ in the dataset $\mathcal{D}$ with $M$ points, the remaining points are sorted by their distance to $\bm{b}$. 
%\end{comment}

In order to visualize the optimization process and the topology of the design spaces, the configurations queried during the optimization process are projected into two-dimensional proximity maps. 
Therefore, we employ Classical Multidimensional Scaling (CMDS).
Multi-dimensional scaling comprises a group of techniques that create a low-dimensional map to visualize the relative positions of a number of objects.
With CMDS, a representation of points $\bm{b} \in \mathcal{D} \subset \mathbb{R}^d$ in terms of two coordinates $\gamma_{1}$ and $\gamma_2$ is found, such that the pairwise distances of the objects in two dimensions in terms of $\gamma$ approximate the true distances under the $d$-dimensional coordinate, i.e. $\Gamma_{ij} \approx D_{ij}$. Here, $\Gamma_{ij}$ and $D_{ij}$ denote the Euclidean distance between $\bm{b}_i$ and $\bm{b}_j$ in 2 and $d$ dimensions, respectively.
% In CMSD feature coordinates $\gamma_{i,j}$ are selected, which optimally preserve the dissimilarity $D_{ij}$ between $N$-dimensional sets of control parameters $\bm{b}_i$ and $\bm{b}_j$. 

% To scan the topology of the design space with respect to local minima, a criterion in $d$ dimensions is monitored~\cite{li2023jfm, wang2023book, WangTY2023}. 
% \begin{figure}[h]
%     \centering
%     \includegraphics[width=0.8\textwidth]{Fig/BO.png}
%     \caption{Sketch of the Bayesian optimization algorithm for design of open-loop controllers for fluid flows \cite{blanchard2021jcp}.}
%    \label{fig:bo}
% \end{figure}
% Bayesian optimization is chosen for its ability to capture prior knowledge about the behavior of black box function and to compromise between exploration and exploitation. We use the likelihood-weighted lower confidence bound (LCB-LW) proposed by Blanchard and Sapsis \cite{blanchard2021jcp} as the output-informed criteria to select the best next point.

% \subsection{Bayesian optimization \cite{blanchard2021jcp, pickering2022ncs}}
% \subsection{Surrogate model}
% \subsection{Sensitivity analysis}

\section{Results}\label{ToC:res}
In the following section, we first analyze the natural jet flow without external forcing. We then describe the optimization process for mixture enhancement in the three aforementioned design spaces -- axisymmetric forcing, $m=0$, superposed axisymmetric and helical forcing, $m\in\{0,1\}$, and axisymmetric actuation combined with two counter-rotating helical modes, $m\in\{-1,0,1\}$. We then analyze the identified optimal excitation patterns. Subsequently, we perform a topological analysis of the highest dimensional design space, based on the combined training data from all optimization runs.

\subsection{Characterization of unforced jet flow}
% --------------------- start figure --------------------
\begin{figure*}[ht]
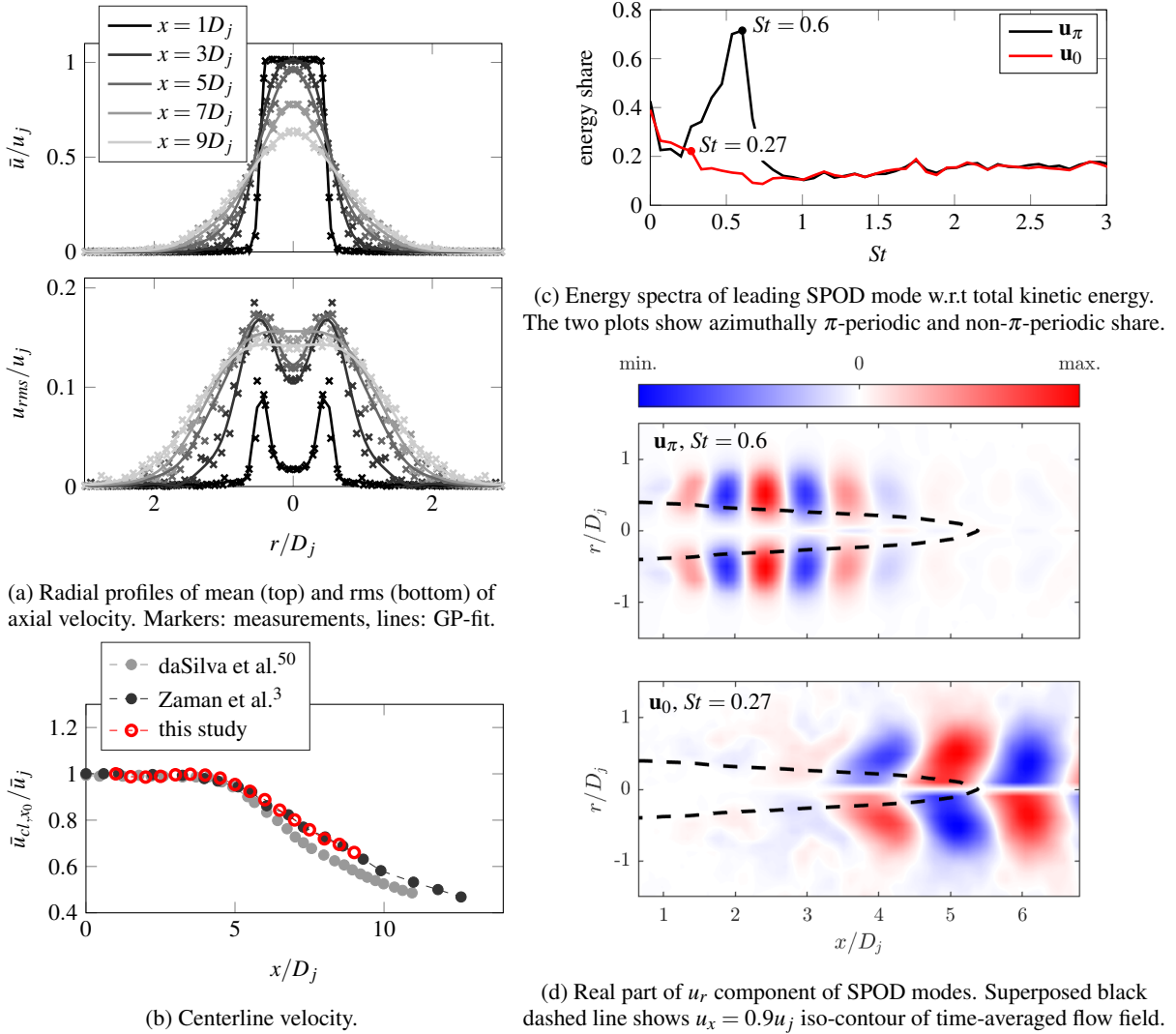

    \centering
     \begin{subfigure}[b]{0.4\textwidth}
         \centering
         \input{Fig/unforced_radial_mean_and_rms}\\
         \caption{Radial profiles of mean (top) and rms (bottom) of axial velocity. Markers: measurements, lines: GP-fit.}
         \label{fig:nat:ur}
         \definecolor{mycolor3}{rgb}{0.2,0.2,0.2}%
\definecolor{mycolor2}{rgb}{0.6,0.6,0.6}%
\definecolor{mycolor1}{rgb}{1,0,0}%
\begin{tikzpicture}

\begin{axis}[%
width=0.8\columnwidth,
height=0.4\columnwidth,
scale only axis,
at={(0,0)},
xmin=0,
xmax=14,
xlabel={$x/D_j$},
ymin=0.4,
ymax=1.3,
ylabel={$\bar{u}_{cl,x_0}/\bar{u}_{j}$},
ylabel near ticks,
legend style={at={(axis cs:0.5,1.1)},anchor=south west, legend cell align=left, align=left, draw=white!15!black}
]
\addplot [color=mycolor2, dashed, mark=*, mark options={solid, mycolor2}]
  table[row sep=crcr]{%
0	0.99355\\
0.45593	0.99\\
1.231	0.99\\
1.77812	0.99355\\
2.18845	0.99\\
2.59878	0.99\\
3.23708	0.98645\\
3.69301	0.98645\\
4.10334	0.98645\\
4.55927	0.96516\\
4.92401	0.94387\\
5.42553	0.90129\\
5.6535	0.87645\\
6.06383	0.83387\\
6.42857	0.79839\\
6.74772	0.7629\\
7.02128	0.72742\\
7.29483	0.70258\\
7.56839	0.67774\\
7.97872	0.64935\\
8.34346	0.62452\\
8.66261	0.60677\\
8.93617	0.58548\\
9.20973	0.56774\\
9.43769	0.55355\\
9.71125	0.53935\\
9.9848	0.52516\\
10.34954	0.51097\\
10.6231	0.49677\\
10.94225	0.48613\\
};
\addlegendentry{daSilva et al.\cite{daSilva2002pf}}

\addplot [color=mycolor3, dashed, mark=*, mark options={solid, mycolor3}]
  table[row sep=crcr]{%
0	1.00065\\
0.59271	1.00065\\
1.09422	1.00065\\
2.23404	0.9971\\
3.23708	0.99355\\
3.96657	0.97935\\
4.42249	0.96871\\
5.10638	0.94387\\
5.56231	0.92258\\
6.06383	0.86226\\
6.79331	0.82323\\
7.29483	0.77\\
8.02432	0.72742\\
8.61702	0.68839\\
9.30091	0.63161\\
9.89362	0.58194\\
10.98784	0.53226\\
11.80851	0.50032\\
12.58359	0.46839\\
};
\addlegendentry{Zaman et al.\cite{zaman1980jfm}}
\addplot [color=mycolor1, dashed, mark=o, mark options={solid, mycolor1}, mark options={scale=1,line width=1.3pt,solid}]
  table[row sep=crcr]{%
0.999975490196079	1\\
1.50004901960784	0.987507720858527\\
1.99997549019608	0.985972568751799\\
2.5	0.989170481386441\\
3	0.99590106811086\\
3.5	0.998340865670933\\
3.99997549019608	0.994665639904184\\
4.5	0.981743501072476\\
5	0.953804382820572\\
5.5	0.924496070084184\\
5.99997549019608	0.888373645505553\\
6.5	0.843341592404265\\
6.99997549019608	0.800938986334077\\
7.5	0.758569839223233\\
7.99997549019608	0.719387017576834\\
8.5	0.695430875963484\\
8.99997549019608	0.660413326665525\\
};
\addlegendentry{this study}
\end{axis}
\end{tikzpicture}%
         \caption{Centerline velocity.}
         \label{fig:nat:ucl}
    \end{subfigure}
    \begin{subfigure}[b]{0.5\textwidth}
        \centering
        \input{Fig/unforced_decomp_SPOD_spectrum}\\
        \caption{Energy spectra of leading SPOD mode w.r.t total kinetic energy. The two plots show azimuthally $\pi$-periodic and non-$\pi$-periodic share.}
        \label{fig:unforced_SPOD_spectrum}
        \centering
        \input{Fig/unforced_decomp_SPOD_modes}
        \caption{Real part of $u_r$ component of SPOD modes. Superposed black dashed line shows $u_x=0.9u_j$ iso-contour of time-averaged flow field.}
        \label{fig:unforced_SPOD_modes}
    \end{subfigure}
    \caption{Characterization of unforced jet flow. Left column: time-averaged characteristics, right column: coherent dynamics.}
    \label{fig:nat}
\end{figure*}
When the acoustic drivers are deactivated, the jet emerges freely into the surrounding air. Figure~\ref{fig:nat:ur} shows radial profiles of the time-averaged and root-mean square (RMS) values of the axial velocity at different downstream locations. The jet, which initially emerges similar to a block profile, expands radially along $x$. While the mean centerline velocity $u_c$ remains approximately constant within the potential core, the shear layer width increases in the direction of flow. Within the shear layers, distinct velocity fluctuations can be seen from the rms plots. 
The length of the potential core $x_p$ is defined to be the axial location $x$ at which $u_{cl}(x) \approx 0.9 u_j$.
After approximately $x_p = 6 D_j$ the shear layers merge, so that the potential core ends and the time-averaged centerline velocity decreases. The centerline velocity is shown in Fig.~\ref{fig:nat:ucl}. For validation purposes, the plot also contains experimental data from Zaman et al.\cite{zaman1980jfm} and numerical data from da Silva et al.\cite{daSilva2002pf}. The velocity profile along the centerline shows high agreement with these studies at similar Reynolds number. At the measuring point used for optimization, $x_0=6.5D_j$, the centerline velocity corresponds to approx. $85\%$ of the exit velocity.

To gain insight into the coherent flow dynamics, the unforced jet is captured using high speed PIV. The acquired time-resolved flow snapshots are then analyzed by means of spectral proper orthogonal decomposition (SPOD), which is a frequency domain form of proper orthogonal decomposition that extracts spatiotemporal coherent modes, oscillating harmonically in time at distinct frequencies~\cite{Towne2018, schmidt_spod_guide}.

In the optimization described below axisymmetric (azimuthal order $m=0$) and helical (azimuthal order $m=\pm1$) excitation schemes are tested. To connect the optimization results with the natural structures in the jet flow, we extract the corresponding flow patterns by decomposing the PIV snapshots. Therefore, the time-resolved flow field in the cross-sectional PIV plane $\mathbf{u}_p = [u_x,u_r]$ is considered, which represent a cut through the three-dimensional flow field $\textbf{u}$,
\begin{equation}
    \mathbf{u}_p(x,r,t) = \mathbf{u}(x,r,\theta=0,t),
\end{equation}
in which $x$ and $r$ range from 0.7 to 6.5$D_j$ and from -1.5 to 1.5$D_j$, respectively.
To extract axisymmetric and non-axisymmetric shares, $\textbf{u}_p$ is decomposed into two parts. 
An azimuthally $\pi$-periodic share $\mathbf{u}_{\pi}$, which is symmetric in $r$, such that $\mathbf{u}_{\pi}(x,r,t) = \mathbf{u}_{\pi}(x,-r,t)$ and a non $\pi$-periodic share $\mathbf{u}_{0}$, which fulfills $\mathbf{u}_{0}(x,r,t) \neq \mathbf{u}_{0}(x,-r,t)$\footnote{Strictly speaking, $\mathbf{u}_{\pi}$ includes all structures of even azimuthal wavenumber in addition to axisymmetric structures, while $\mathbf{u}_{0}$ covers all odd azimuthal wavenumbers.}. The decomposition is applied to every PIV snapshot as follows:
\begin{align}
    \mathbf{u}_{\pi}(x,r,t) &= \frac{1}{2}(\mathbf{u}_p(x,r,t) + \mathbf{u}_p(x,-r,t)),\\
    \mathbf{u}_{0}(x,r,t) &= \frac{1}{2}(\mathbf{u}_p(x,r,t) - \mathbf{u}_p(x,-r,t)).
\end{align}
Note that this decompostion is comprehensive because $\mathbf{u}_{0} + \mathbf{u}_{\pi}$ recovers the original PIV snapshots $\textbf{u}_p$. Applying SPOD to the decomposed snapshots yields the spectra shown in Fig.~\ref{fig:unforced_SPOD_spectrum}. The plot shows the share of the most energetic SPOD mode relative to the total kinetic energy at different frequencies. The spectrum of the symmetric ($\pi$-periodic) flow component shows a clear peak at $St\approx 0.6$. In this frequency bin the leading mode comprises about 70\% of the kinetic energy which indicates a strong coherent structure with distinct oscillation frequency. The spectrum of the asymmetric (non $\pi$-periodic) component, on the other hand, does not show any distinct frequency peaks. However, the highest energy share of the leading mode can be found below St = 0.3, which is potentially harmonically related to the axisymmetric distinct one. In this frequency range, the energy share of over 20\% is distributed over a wider frequency range. The associated SPOD modeshapes are shown in Fig.~\ref{fig:unforced_SPOD_modes}. The colored background depicts the real part of the radial component of the modes. For visual guidance, an iso-contour of the mean flow field at $\bar{u}_x = 0.9u_j$ is superposed to the fields, indicating the potential core. The $\mathbf{u}_\pi$ mode shows a symmetric, streamwise periodic pattern with vortices within the jet shear layer that oscillate mainly along the potential core in the range between $x/D_j=1$ and 4.5. At the end of the potential core, the coherent oscillation at this frequency appears to have dissipated. The shown $\mathbf{u}_0$ mode at $St=0.27$, on the other hand, shows a helical pattern around the jet axis, which oscillates most strongly in the range $x/D_j \geq 4$. The wavelength of the vortex structure is about twice as long as in the $\mathbf{u}_0$ mode, which indicates a similar convection speed of the vortices of both structures, given the roughly half frequency of the $\mathbf{u}_0$ mode. The coherent motion in the flow is clearly distinguishable until the downstream end of the PIV domain.

\subsection{Optimization processes}
\begin{figure*}[ht]
\begin{subfigure}[t]{0.33\textwidth}
    \centering
    \includegraphics[width=\columnwidth]{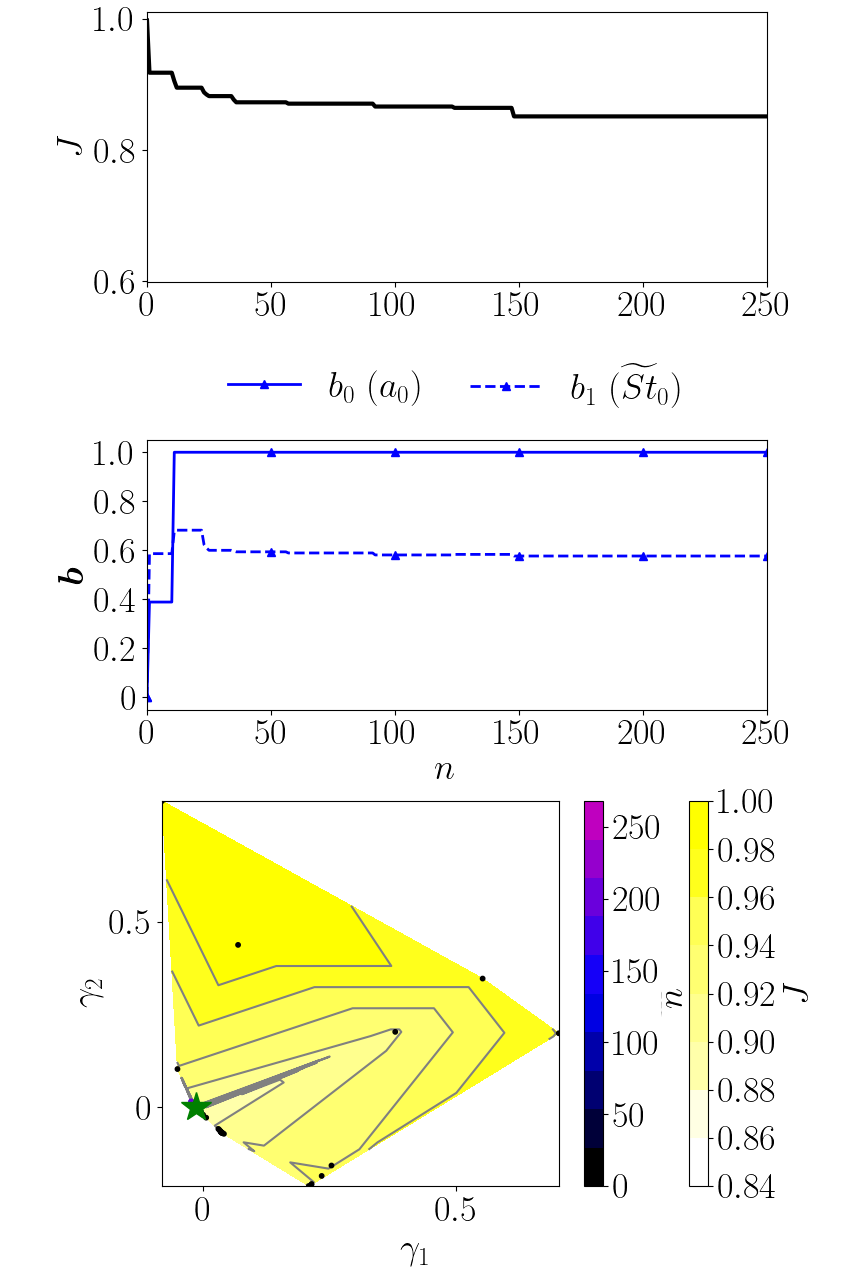}
    % \rule{7cm}{3cm}
   \caption{Axissymetric forcing, $\bm{b} \in \bm{B}_1.$}
    \label{fig:results:learning2d}
   \end{subfigure}
   \begin{subfigure}[t]{0.33\textwidth}
    \includegraphics[width=\columnwidth]{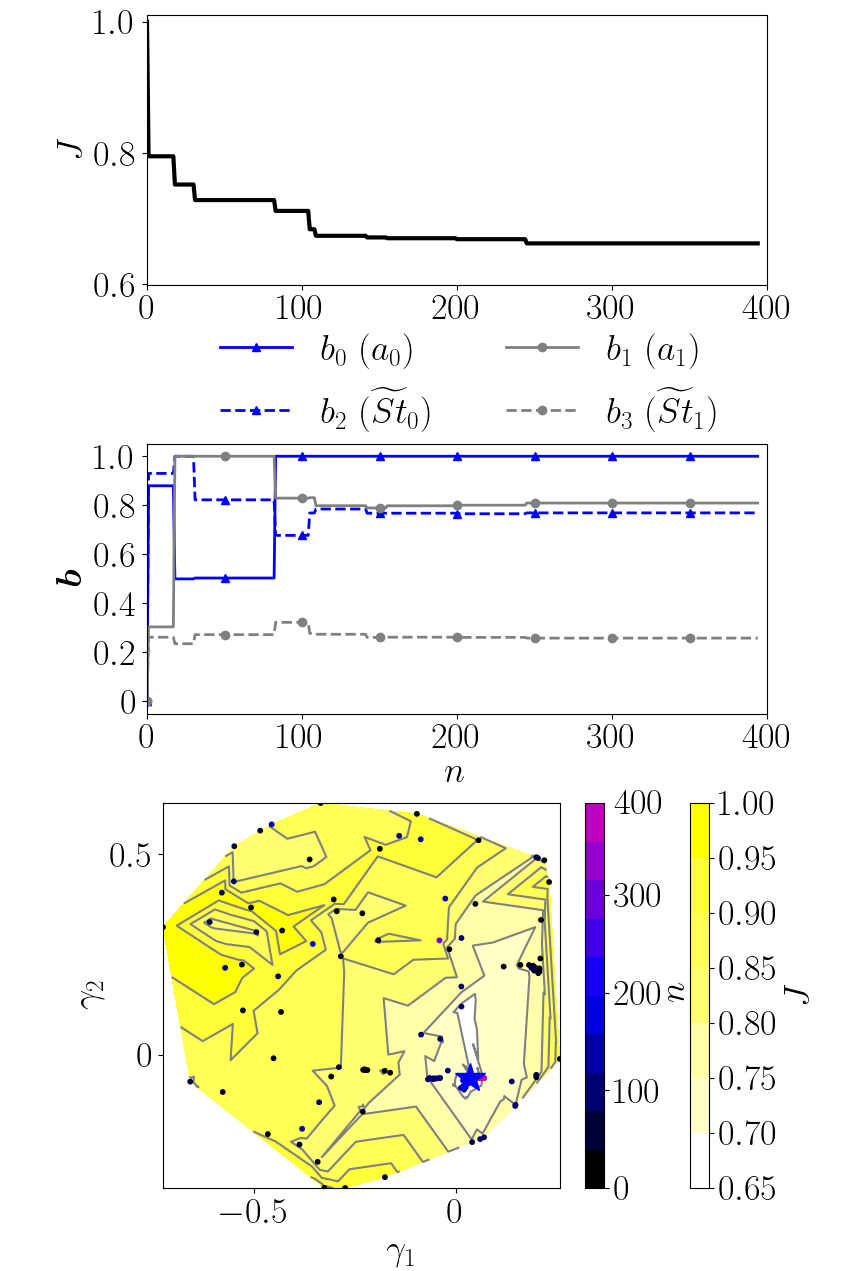}
    \caption{Axissymetric and helical forcing, $\bm{b} \in \bm{B}_2.$}
    \label{fig:results:learning4d}
   \end{subfigure}
   \begin{subfigure}[t]{0.33\textwidth}
    \includegraphics[width=\columnwidth]{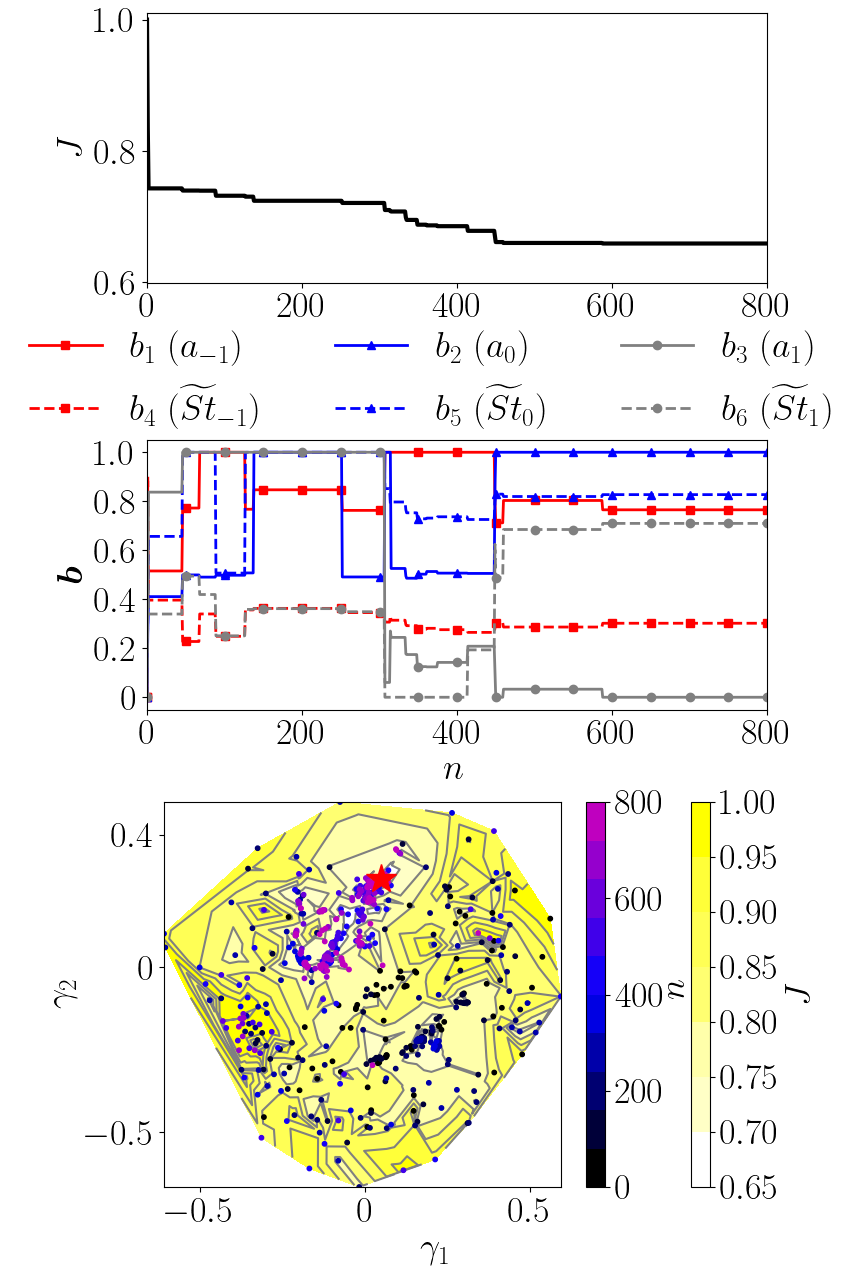}
    \caption{Axissymetric, helical and counter-rotating helical forcing, $\bm{b} \in \bm{B}_3.$}
    \label{fig:results:learning6d}
   \end{subfigure}
    \caption{Learning process in the three search spaces. Top row: variation of the minimal objective values (learning curves). Center row: Evolution of the input parameters tested during the optimization. Bottom row: Proximity map of the tested samples and approximated search spaces.}
    \label{fig:results:learning_all}
\end{figure*}
\label{ToC:res:opt}
After having identified the naturally occurring coherent structures in the jet flow, the optimization process of the flow excitation, Eq.~\ref{eq:opti_problem} is described in the following. We therefore gradually increase the complexity of the optimization problem by
considering the following three search spaces of dimension $2,4$ and $6$:
\begin{enumerate}
    \item axisymmetric forcing, $\bm{b}\in \bm{B}_1 \subset \mathbb{R}^2,~m=0$,
    \item superposed axisymmetric and helical forcing\\ $\bm{b}\in \bm{B}_2 \subset \mathbb{R}^4,~m\in\{0, 1 \}$,
    \item superposed axisymmetric, helical and counter-rotating helical forcing $\bm{b}\in \bm{B}_3 \subset \mathbb{R}^6,~m\in\{-1, 0, 1 \}$.
\end{enumerate}
Figure \ref{fig:results:learning_all} shows the optimization processes in the above search spaces in terms of (i) the learning curves, i.e. the cost value $J$ corresponding to the current minimum after $m$ evaluations, (ii) the variation of the tested optimization parameters $\bm{b}_m$, and (iii) proximity maps of the search spaces that show the queried control parameters as markers, colored by the iteration sequence $m$ on a background that indicates the optimization function, interpolated and colored by the cost functional $J$ of all samples. The lightest areas are related to optimal mixing.
\subsubsection{Axisymmetric forcing}
The process of optimizing the actuation parameters for axisymmetric forcing with iteration $n$ is shown in Fig.~\ref{fig:results:learning2d}. To find an optimum in this search space $\bm{B}_1 \subset \mathbb{R}^2$, 
$N_{max}=250$ optimization iterations are performed. Within the first $20$ iterations, larger variations in the parameters are recognizable, associated with an exploitative parameter search across the input space. Subsequently, the tested configurations cluster in one parameter region. For later iterations, the optimization scheme performs finer corrections. After 250 iterations, the minimal $J$ value is $0.85$, with the optimized solution $\bm{b}=(1, 0.57)$. The parameter ${b}_1$ which represents the actuation amplitude converges to the upper limit of the search range $a_0 = 1$ (triangle-solid line in center row of Fig.~\ref{fig:results:learning2d}).
The results indicates that the mixing effect increases with the amplitude of the axisymmetric forcing. This is also validated by the proximity map where the samples converge to the border of the map.
The optimized Strouhal number corresponding to ${b}_2$, $St_0$ is equal to $0.61$ (triangle-dashed line). The interpolated control landscape mainly indicates a large basin of attraction around a global minimum, which leads to rapid convergence of the BO routine.
\subsubsection{Combined axisymmetric and helical forcing}
With one axisymmetric mode ($m=0$) and one helical mode ($m=1$) superposed, the cost function $J$ can be minimized to $0.66$ as shown in Fig.~\ref{fig:results:learning4d}. Due to the increased dimension and the resulting more complex optimization landscape, the algorithm uses approximately the first 120 iterations for larger parameter steps in which the search space is explored. In the subsequent steps, finer parameter adjustments can be recognized and the samples converge to a large basin of attraction of the proximity map as the scattered markers in the proximity map in Fig.~\ref{fig:results:learning4d} show. The optimized actuation parameters $\bm{b}$ are found to be $(1, 0.81, 0.77, 0.26)$ after $N_{max} = 400$ evaluations.
The optimal forcing thus consists of an axisymmetric forcing at maximum amplitude ($a_0=1$, triangle-solid line in Fig.~\ref{fig:results:learning4d}), and a helical mode with an amplitude below the maximum ($a_1=0.77$, dot-solid line). The identified optimal Strouhal numbers are $St_0=0.79$ (triangle-dashed line), and $St_1=0.33$ (dot-solid line). The proximity map in Fig.~\ref{fig:results:learning4d} again indicates a control landscape that mainly consists of a large area of attraction around the found minimum.

\subsubsection{Combined axisymmetric, helical and counter-rotating helical forcing}
Figure~\ref{fig:results:learning6d} shows the optimization in $\bm{B}_3$, which combines superposed forcing of one axisymmetric mode ($m=0$), one helical mode  ($m=1$), and another counter-rotating helical mode ($m=-1$). This 6-dimensional parameter space is optimized within $N_{max}=800$ iterations. The parameter variations shown in the center plot of Fig.~\ref{fig:results:learning6d} show several phases of the optimization process. For $n<150$, large variations again indicate an explorative sampling strategy from BO, which can also be seen in the proximity map in Fig.~\ref{fig:results:learning6d}. During iterations $n=150$ to 250, the parameter curves show only small changes and the proximity map reveals that the optimizer searches within a basin in the search space. During this phase, configurations with $St_1 \approx St_{-1}$ are tested. After around 300 iterations, the optimizer changes into an exploration mode again before after $n>450$ iterations, samples are tested in the vicinity of the final minimum. The final optimum $\bm{b} = (0.77, 1, 0, 0.3, 0.83, 0.71)$ is similar to the optimum in the two-mode search space described above with $J=0.65$. With the amplitude of one helical mode $a_{1}$ set to zero
the optimal forcing is formed by a two-mode pattern with a axisymmetric and a helical contribution.
The axisymmetric and helical mode are enforced at amplitudes $a_{0}=1$, $a_{-1}=0.77$, and frequencies $St_{0}=0.85$, $St_{-1}=0.37$, respectively. The proximity map shows that this minimum is located in a second large basin of attraction,
where the most samples are queried and which is deeper than the suboptimal one tested during $n=150$ to 250.
% % --------------------- start figure ---------------------
% \begin{figure}
%     \centering
%     \includegraphics[width=\linewidth]{Fig/learningmap_min_woj.png}
%     % \rule{7cm}{3cm}
%     \caption{Proximity map of the 3-mode search spaces with the minima derived from the discrete data.
%     }
%    \label{fig:results:learning_minima}
% \end{figure}
% % --------------------- end figure ---------------------

\subsection{Topological analysis and the optimal excitation patterns}
% --------------------- start figure ---------------------
\begin{figure}
    \centering
    \includegraphics[width=\linewidth]{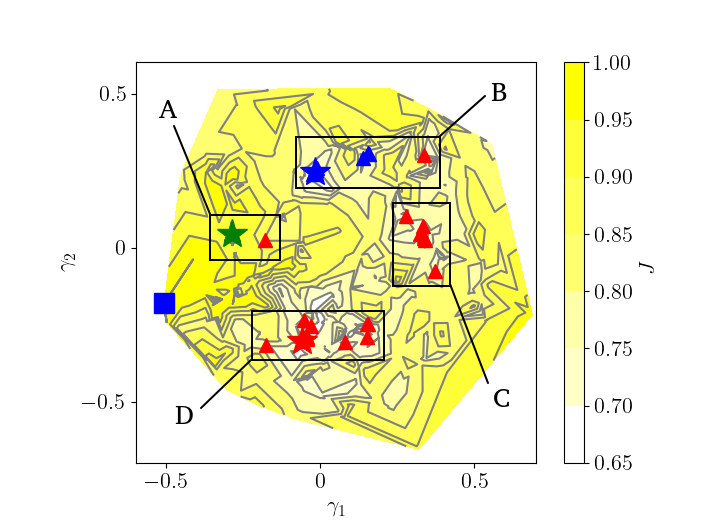}
    % \rule{7cm}{3cm}
    \caption{Proximity map of the solutions in search spaces with the minima derived from discrete data. 
    Square --- unforced case.
    Green symbols --- $\bm{b} \in \bm{B}_1$, %axisymmetric forcing $m=0$, 
    blue symbols --- $\bm{b} \in \bm{B}_2$, %combined axisymmetric and helical forcing $m=0,1$ 
    and red symbols --- $\bm{b} \in \bm{B}_3$.%combined axisymmetric forcing, helical and counter-rotating helical forcing $m=-1,0,1$.
    Triangles --- local minima. Stars --- optimal solutions in $\bm{B}_1$, $\bm{B}_2$, and $\bm{B}_3$.
    }
   \label{fig:results:learning_minima}
\end{figure}
% --------------------- end figure ---------------------
As shown in the proximity maps of optimizations in $\bm{B}_1$, $\bm{B}_2$, and $\bm{B}_3$ (bottom of Figs.~\ref{fig:results:learning2d}, \ref{fig:results:learning4d}, and \ref{fig:results:learning6d}), the search space topology becomes more complex as the dimension increases.
The optimal solution also increases in complexity from $\bm{B}_1$ to $\bm{B}_2$. However, the optimizations in $\bm{B}_2$ and $\bm{B}_3$ converge to a similar solutions (center of Figs.~\ref{fig:results:learning2d}, \ref{fig:results:learning4d}, and \ref{fig:results:learning6d}). This motivates the examination of the individual optima in relation to each other.
% Before we analyze the detected optima and the associated jet response in detail, we examine their relation to each other. 
% Therefore, we make use of the fact that the three search spaces are subsets of each other, $\bm{B}_1 \subset \bm{B}_2 \subset \bm{B}_3$. 
% -------------- Begin table
\begin{table}[b]
    \centering
    \begin{tabularx}{\columnwidth}{@{}cYYYYYY@{}}
    \toprule
    $J$ & ${b}_1, a_{-1}$ & ${b}_2,a_{0}$ & ${b}_3,a_{1}$ & ${b}_4,\widetilde{St}_{-1}$ & ${b}_5,\widetilde{St}_{0}$     & ${b}_6,\widetilde{St}_{1}$\\
    \midrule    
    $0.87$           & $0$  & $1$  & $0$  & $0$  &  $0.59$  & $0$\\
    % $0.87$           & $0$  & $1$  & $0$  & $0$  &  $0.58$  & $0$\\
    % $0.87$           & $0$  & $1$  & $0$  & $0$  &  $0.58$  & $0$\\
    $0.86$           & $0$  & $1$  & $0$  & $0$  &  $0.58$  & $0$\\
    $0.86$           & $0$  & $1$  & $0$  & $0$  &  $0.56$  & $0$\\
    $0.85$ ($J_{min}$)          & $0$  & $1$  & $0$  & $0$  &  $0.57$  & $0$\\    
    $0.85$ (in $\bm{B}_3$)   & $0.04$  &$0.92$ &$0.19$ &$0.33$ &$0.53$ &$0.05$\\
    \bottomrule
    \end{tabularx}
    \caption{The input parameters of the minima located in Region A (axisymmetric forcing).}
    \label{tab:minima_a}
\end{table}
Therefore, a topological data analysis is performed on the data from all three optimizations using persistent data topology as described in \S~\ref{ToC:PDT}. The input vectors from $\bm{B}_1 \subset \mathbb{R}^2$ and $\bm{B}_2 \subset \mathbb{R}^4$ are augmented to $6$ dimensions by adding $0$ for the additional entries.
This allows us to combine the queried forcing configurations from all three optimizations to form a large data set consisting of 1450 samples in $\bm{B}_3$. This dataset is used to construct a two-dimensional proximity map of the search space, shown in Fig.~\ref{fig:results:learning_minima}. The background is colored by the cost $J$ and the symbols denote the optima found in space $\bm{B}_1$ (green), $\bm{B}_2$ (blue) and $\bm{B}_3$ (red). Here, the stars highlight the best solutions, and the triangles denote local minima derived from the enriched dataset.
The derived minima include $4$ solutions within the data from $\bm{B}_1$, $5$ from the search in $\bm{B}_2$, and $16$ in $\bm{B}_3$.
These minima are distributed in four basins, named A, B, C, D as grouped by the rectangles in Fig.~\ref{fig:results:learning_minima}. 
% In the figure, the previously found minima are marked with stars. The colors indicate the results from the three optimizations in search. In addition, local minima are indicated with triangles in the map, colored accordingly.
We quantify the depth of these basins by the cost difference of the minima inside, $\delta J$.
The Basin D features the largest depth $\delta J=0.06$, followed by the Basin B, C, and A with $\delta J=0.04$, $\delta J=0.03$, and $\delta J=0.02$ respectively.
The topological analysis shows that the optimal axisymmetric excitation only forms a local plateau (Region A in Fig.~\ref{fig:results:learning_all}) in $\bm{B}_3$, where all local minima identified from search in $\bm{B}_1$ overlap. 
Although higher objective values can be seen in the vicinity, the objective value in the majority of the proximity map is lower than the first optimum (0.85), which indicates a strong benefit from incorporating helical modes.
The Basins B and D are aligned approximately symmetrical with respect to $\gamma_2=0$ and include the optimized solutions from $\bm{B}_2$ and $\bm{B}_3$, respectively both at $\gamma_1\approx0$ (blue and red star).
This symmetry is related to the fact that the optimization in $\bm{B}_3$ identifies a similar forcing pattern as the one found in $\bm{B}_2$ but with a helical mode of order $m=-1$ instead of $m=1$. Accordingly, the associated cost values are similar.
% The following text still needs to be updated: 24.3.2024
% The depth of the basin on the top left which contains the global minimum is $\delta J=0.03$, and that on the right bottom is more shallow with $\delta J=0.01$.

\begin{figure*}[t!]
    \centering
    \includegraphics[width=0.8\textwidth]{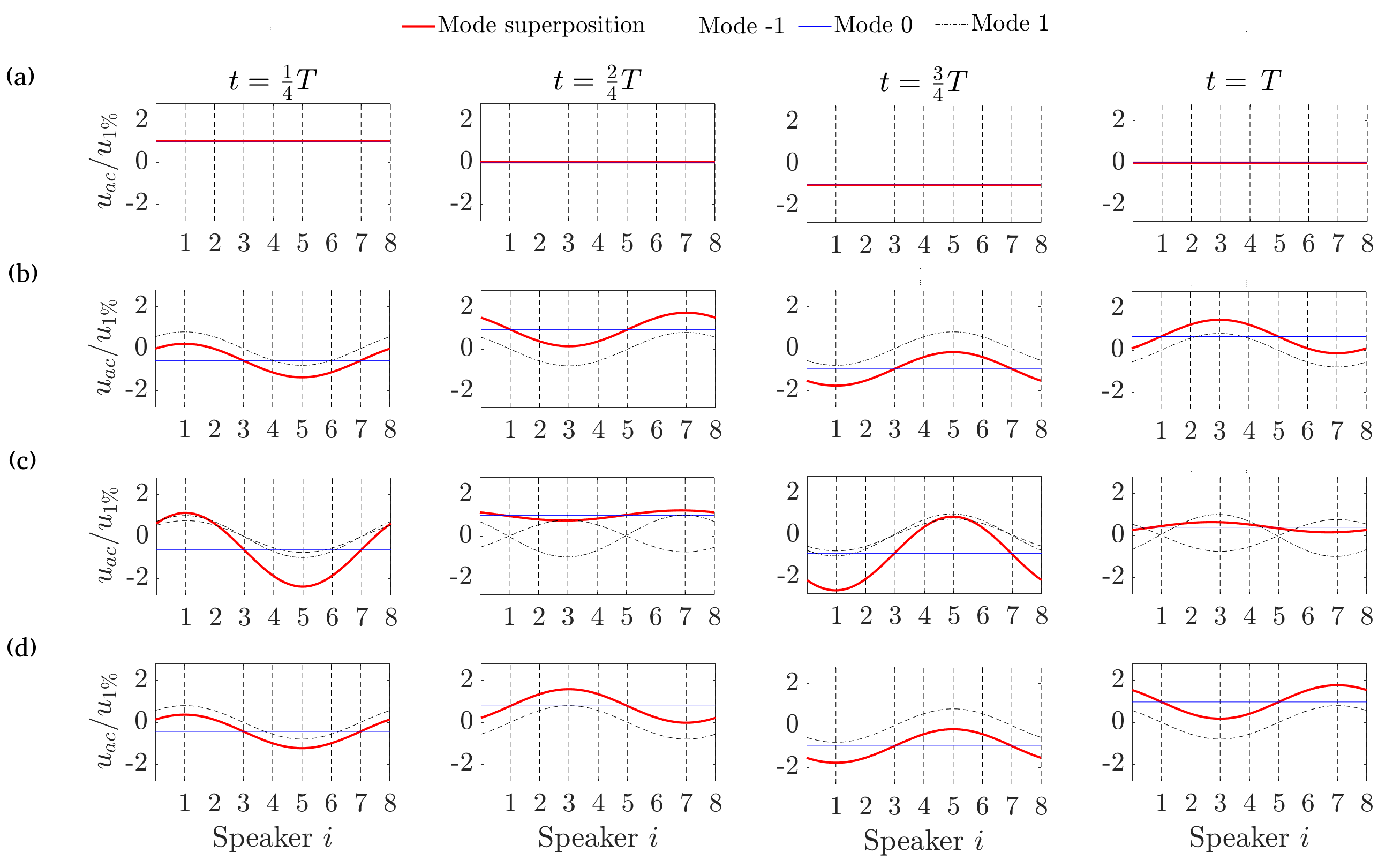}
    \caption{The actuation signals in one period of 
            (a) the optimal axisymmetric forcing in basin A, 
            (b) the optimal axisymmetric-helical forcing in basin B, 
            (c) the optimal axisymmetric-flapping forcing in basin C, 
            and (d) the optimal axisymmetric-helical forcing in basin D.
            % the optimized non-axisymmetric solution (b) in 2-mode forcing spaces,
            % and the optimized non-axisymmetric solution (c) in 2-mode forcing spaces.
            }
   \label{fig:results:act}
\end{figure*}

% -------------- End table
% -------------- Begin table
\begin{table}
    \centering
    \begin{tabularx}{\columnwidth}{@{}cYYYYYY@{}}
    \toprule
    $J$ & ${b}_1, a_{-1}$ & ${b}_2,a_{0}$ & ${b}_3,a_{1}$ & ${b}_4,\widetilde{St}_{-1}$ & ${b}_5,\widetilde{St}_{0}$     & ${b}_6,\widetilde{St}_{1}$\\
    \midrule    
    % $0.67$          & $0$     & $1$  & $0.79$  & $0$  &  $0.76$  & $0.26$\\
    % $0.67$          & $0$     & $1$  & $0.80$  & $0$  &  $0.76$  & $0.26$\\
    % $0.67$          & $0$     & $1$  & $0.80$  & $0$  &  $0.76$  & $0.26$\\    
    $0.67$          & $0$     & $1$  & $0.79$  & $0$  &  $0.76$  & $0.26$\\
    $0.67$          & $0$     & $1$  & $0.79$  & $0$  &  $0.78$  & $0.27$\\
    $0.67$          & $0$     & $1$  & $0.78$  & $0$  &  $0.76$  & $0.26$\\
    $0.69$          & $0$     & $1$  & $0.80$  & $0$  &  $0.77$  & $0.26$\\    
    $0.67$ ($J_{min}$)          & $0$     & $1$  & $0.80$  & $0$  &  $0.76$  & $0.26$\\
    $0.71$ (in $\bm{B}_3$)     & $0.13$  & $1$  & $1$  & $1$  &  $0.71$  & $0.32$\\
    \bottomrule
    \end{tabularx}
    \caption{The input parameters of the minima located in Region B (axisymmetric-helical forcing).}
    \label{tab:minima_b}
\end{table}
% -------------- End table
% -------------- Begin table
\begin{table}
    \centering
    \begin{tabularx}{\columnwidth}{@{}cYYYYYY@{}}
    \toprule
    $J$ & ${b}_1, a_{-1}$ & ${b}_2,a_{0}$ & ${b}_3,a_{1}$ & ${b}_4,\widetilde{St}_{-1}$ & ${b}_5,\widetilde{St}_{0}$     & ${b}_6,\widetilde{St}_{1}$\\
    \midrule
    $0.73$ & $1$    & $0.50$  & $1$ & $0.25$ & $0.51$ & $0.25$\\
    $0.74$ & $0.85$ & $1$     & $1$ &  $0.36$ & $1$    &  $0.36$\\
    $0.73$ & $0.78$ & $0.51$  & $1$ &  $0.37$ & $1$    &   $0.34$\\
    $0.72$ & $0.71$ & $0.49$  & $1$ &  $0.35$ & $1$    &  $0.37$\\
    $0.71$ ($J_{min}$) & $0.76$ & $0.49$  & $1$ &  $0.35$ & $1$    &   $0.35$\\
    \bottomrule
    \end{tabularx}
    \caption{The input parameters of the minima located in Region C (axisymmetric-flapping forcing).}
    \label{tab:minima_c}
\end{table}
% -------------- End table
% -------------- Begin table
\begin{table}
    \centering
    \begin{tabularx}{\columnwidth}{@{}cYYYYYY@{}}
    \toprule
    $J$ & ${b}_1, a_{-1}$ & ${b}_2,a_{0}$ & ${b}_3,a_{1}$ & ${b}_4,\widetilde{St}_{-1}$ & ${b}_5,\widetilde{St}_{0}$     & ${b}_6,\widetilde{St}_{1}$\\
    \midrule
    $0.68$ & $1$     & $0.51$ &$0.14$ &$0.28$ &$0.74$ &$0$   \\
    $0.67$ & $1$     & $0.51$ &$0.21$ &$0.27$ &$0.73$ &$0.19$\\
    $0.68$ & $1$     & $0.51$ &$0.21$ &$0.26$ &$0.73$ &$0.22$ \\
    $0.67$ & $1$     & $0.64$ &$0.07$ &$0.28$ &$0.79$ &$0.48$\\
    $0.66$ & $0.71$ & $1$     &$0$     &$0.3$ &$0.83$ &$0.49$\\
    $0.66$ & $0.8$ & $1$     &$0.03$ &$0.29$ &$0.82$ &$0.69$\\
    $0.66$ & $0.8$ & $0.97$ &$0$     &$0.29$ &$0.82$ &$0.46$\\
    $0.71$ & $0.81$ & $1$     &$0$     &$0.3$ &$0.82$ &$0.6$\\
    $0.65$ ($J_{min}$) & $0.77$ & $1$     &$0$     &$0.3$ &$0.83$ &$0.71$ \\
    \bottomrule
    \end{tabularx}
    \caption{The input parameters of the minima located in Region D (axisymmetric-helical forcing).}
    \label{tab:minima_d}
\end{table}
% -------------- End table
As a close examination of the minima in the four regions reveals, the four basins can be classified into distinct forcing categories according to the forcing patterns related to the contained minima. The forcing parameters of the minima are grouped and listed in Tabs.~\ref{tab:minima_a}, \ref{tab:minima_b}, \ref{tab:minima_c} and \ref{tab:minima_d}. In the following, we identify a characteristic forcing pattern for each of the regions. Figure~\ref{fig:results:act} shows how the four forcing patterns evolve in time. The figure shows four snapshots in one period $T$, where $T$ is computed from the lowest frequency contained in the actuation. The velocity variation related to the azimuthal location along the nozzle edge is shown, which is applied by the eight loudspeakers the locations of which are indicated by grey vertical lines. The red solid lines denote the actuation signals, which are formed as the superposition of the $m=-1$ (dashed line), $m=0$ (solid line), and $m=1$ (dash-dotted lines) modes.

% --------------------------- begin fig ---------------------------------
% --------------------------- end fig ---------------------------------
% Region A:
Table \ref{tab:minima_a} lists all minima in Basin A. The table shows that they can all be classified as axisymmetric forcing with a (near) maximum amplitude at a Strouhal number of around $0.61$ (denormalized $\bm{b}_5 = 0.57$), and amplitudes for $m=-1$ and $m=1$ close to $0$, including one minimum in $\bm{B}_3$. A characteristic actuation signal for Region A for each loudspeaker $i$ thus reads
\begin{equation}
    \frac{u_{ac,i}^A(t)}{u_{1\!\%}} = \sin\Big(0.61 \omega_c t \Big).
\label{Eq:command:2d}
\end{equation}
The top row of Fig.~\ref{fig:results:act} shows the corresponding actuation commands evolving with time. As shown in the figure, the instantaneous amplitudes of all actuators are equal in every instance in time, forming the axisymmetric actuation.

% Region B:
All minima identified in Region B are listed in Tab.~\ref{tab:minima_b}. The configurations feature a (near) zero amplitude for azimuthal order $-1$ and a maximum amplitude for $m=0$ with Strouhal number $St_0=0.79$ ($\bm{b}_5 = 0.77$). The helical mode of order $m=1$ is included with an amplitude of $a_1=0.8$ and a frequency of $St_1=0.33$ ($\bm{b}_6 = 0.26$), forming the following characteristic velocity signal of Region B
\begin{equation}
\begin{aligned}
    \frac{u_{ac,i}^B(t)}{u_{1\!\%}} &= \sin\Big(0.79 \omega_c t \Big)+ 0.8~ \sin\Big(0.33 \omega_c t + \frac{i \pi}{4}\Big).
    % a_i(t) & = 0.07 \sin\left(2 \pi \times 45.56 t \right) \\
    % & ~ + 0.04 \sin\left(2 \pi \times 19.10 t + \frac{2i \pi}{8}\right).
\label{Eq:command:4d}
\end{aligned}
\end{equation}
The frequency ratio between the $m=0$ mode and $m=-1$ mode is $2.39$.
The signal (red line in Fig.~\ref{fig:results:act}b) thus consists of a axisymmetric mode ($m=0$, solid blue line) and a circumferentially traveling $m=-1$ mode (dash-dotted blue line).

% Region C:
As presented in Tab.~\ref{tab:minima_c}, the minima in Basin C have in common that $St_1 \approx St_{-1}$ at high amplitudes, with slight asymmetry in amplitudes. As shown in Fig.~\ref{fig:results:act}c, the two spinning modes with equal frequency combine to a standing mode of azimuthal order 1 which generates a flapping motion around a nodal line between speakers $3$ and $7$. In most of the forcings in Basin C, this is superposed with an axisymmetric mode at $St_0 = 1$ and an amplitude pf $a_0\approx0.49$. Accordingly, this basin is found to be associated to axisymmetric-flapping actuation. The associated axisymmetric-flapping velocity forcing reads 
\begin{equation}
\begin{aligned}
        \frac{u_{ac,i}^C(t)}{u_{1\!\%}} &= 0.49~\sin\Big(\omega_c t \Big)
        + 0.76~\sin\Big( 0.41 \omega_c t - \frac{i \pi}{4}\Big)\\
        &+  \sin\Big( 0.41 \omega_c t + \frac{i \pi}{4}\Big).
\label{Eq:command:6d flap}
\end{aligned}
\end{equation}

%Region D:
In Tab.~\ref{tab:minima_d}, the input parameters of the nine local minima identified in the deepest basin are given. The dominating modes are a helical mode with $m=-1$ and an axisymmetric mode $m=0$, leaving the amplitude of the other helical mode $a_1$ at values below $0.21$. The frequencies of the $m=0$ and $m=-1$ modes are around $St=0.85$ and $St=0.37$ ($\bm{b}_5=0.83$ and $\bm{b}_4=0.3$) and are similar for all minima such that the basin can be clearly associate to a characteristic axisymmetric-helical forcing scheme. The frequency ratio between axisymmetric and helical mode is 2.29, thus similar to the one from Basin B. The associated forcing command reads
\begin{equation}
\begin{aligned}
        \frac{u_{ac,i}^D(t)}{u_{1\!\%}} &= \sin\Big(0.85 \omega_c t\Big) + 0.8~\sin\Big(0.37 \omega_c t - \frac{i \pi}{4}\Big).
    % + 0.0\sin\left(259 t + \frac{2i \pi}{8}\right)
    % a_i(t) & = 0.07\sin\left(2 \pi \times 48.58 t\right)\\
    % & ~ + 0.04\sin\left(2 \pi \times 21.39 t - \frac{2i \pi}{8}\right).
    % % + 0.0\sin\left(259 t + \frac{2i \pi}{8}\right)
\end{aligned}
\label{Eq:command:6d}
\end{equation}
The actuation pattern is shown in Fig.~\ref{fig:results:act}d. The amplitude distribution comprises a azimuthal mode and a spinning motion which rotates in an opposite direction to Fig.~\ref{fig:results:act}b.

In summary, the topological analysis has identified four characteristic forcings, three of which are found to be optimal (A,B,D) and a sub-optimal one (C). We will continue with an analysis of the jet response to the optimal patterns. Due to the similarities between forcings in Region B and D, we group them and only investigate the optimal axisymmetric (A) and the optimal axisymmetric-helical (B/D) forcing in the following.

\subsection{Jet response to optimal excitation patterns}
In this section, the jet response to the identified optimal forcing patterns (axisym. $m=0$: Eq.\eqref{Eq:command:2d} and axisym. + helical $m\in\{-1,0\}$ Eq.~\eqref{Eq:command:6d}) are analyzed by means of PIV acquisition. % and helical + counter-rot. helical $m=\{-1,1\}$ Eq.\eqref{Eq:command:6d flap}) are analyzed in detail. 
We first describe the induced formation of vortices, then investigate the influence of the actuation on the time-averaged flow statistics before examining the flow dynamics by means of SPOD.
% --------------------------- vorticity snapshots ---------------------------------
% --------------------------- begin fig ---------------------------------
\begin{figure}
    \centering
    \includegraphics[width=0.8\linewidth]{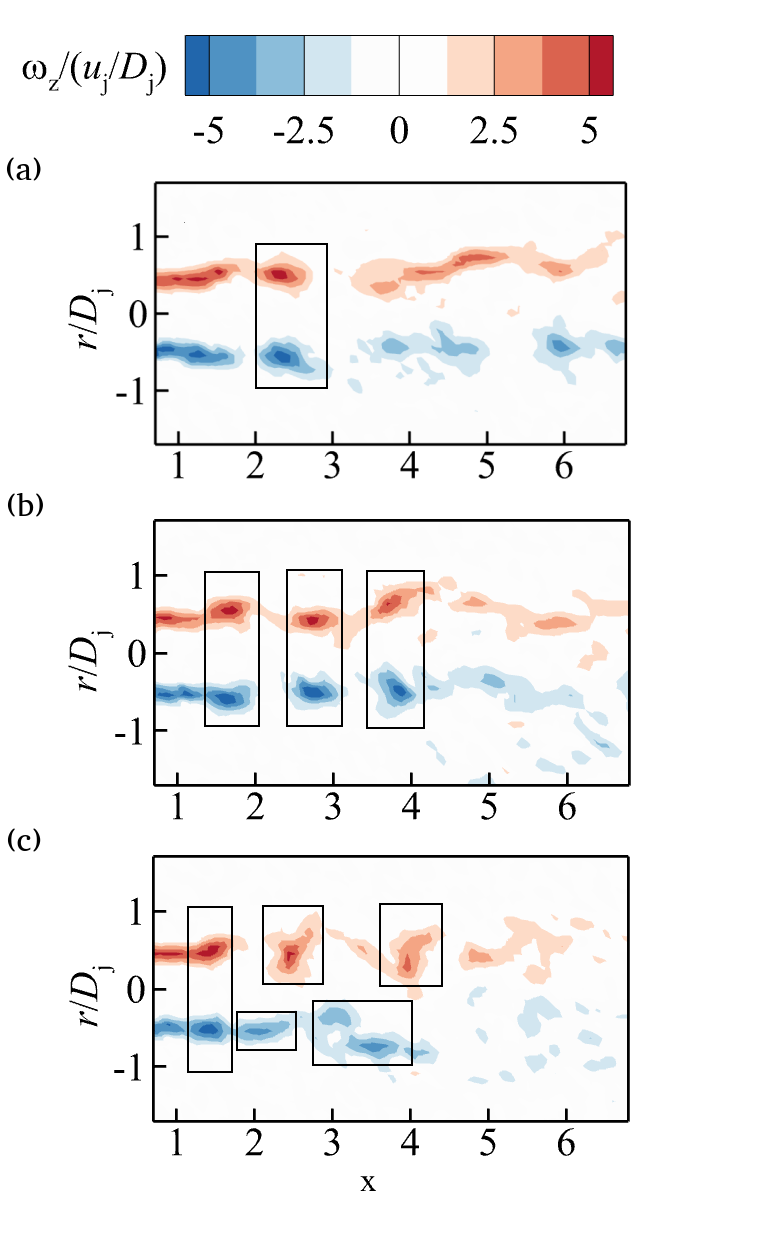}
    \caption{Instantenous vorticity fields of the unforced (a) and the forced cases with the optimized axisymmetric (b) and non-axisymmetric actuation (c).}
    \label{fig:results:flow_vort}
\end{figure}
% --------------------------- end fig ---------------------------------
Figure~\ref{fig:results:flow_vort} shows exemplary instantaneous snapshots of vorticity without and with different forcings in the PIV plane. Without forcing, the shear layers roll up and form a mostly axisymmetric vorticity distribution. Vortex rings are formed that appear as counter-rotating pairs of vortices in the acquisition plane. The first distinct vortex rings are recognizable at $x=2.5D_j$, as exemplified in Fig.~\ref{fig:results:flow_vort}a.
When the optimal axisymmetric forcing is applied (Fig.~\ref{fig:results:flow_vort}b), vortex generation is observed closer to the nozzle, around $x=1.5D_j$. More pairs of symmetric vortices are visible in the near field, potentially due to the slightly higher Strouhal number of the forcing $St=0.61$, compared to the natural frequency of $St=0.58$ or due to stronger induced vorticity, which is thus easier to distinguish. The symmetry of vortex generation is interrupted when a helical component ($m=-1$) is added to the forcing as shown in Fig.~\ref{fig:results:flow_vort}c. The symmetric vorticity distribution is visible up to $x\approx1.5D_j$. Around $x=2D$, vortices alternately detach from the upper and lower section of the shear layer, indicating a helical vortex formation around the jet axis.

% --------------------------- mean fields ---------------------------------
% --------------------------- begin fig ---------------------------------
\begin{figure}
    \centering
    \input{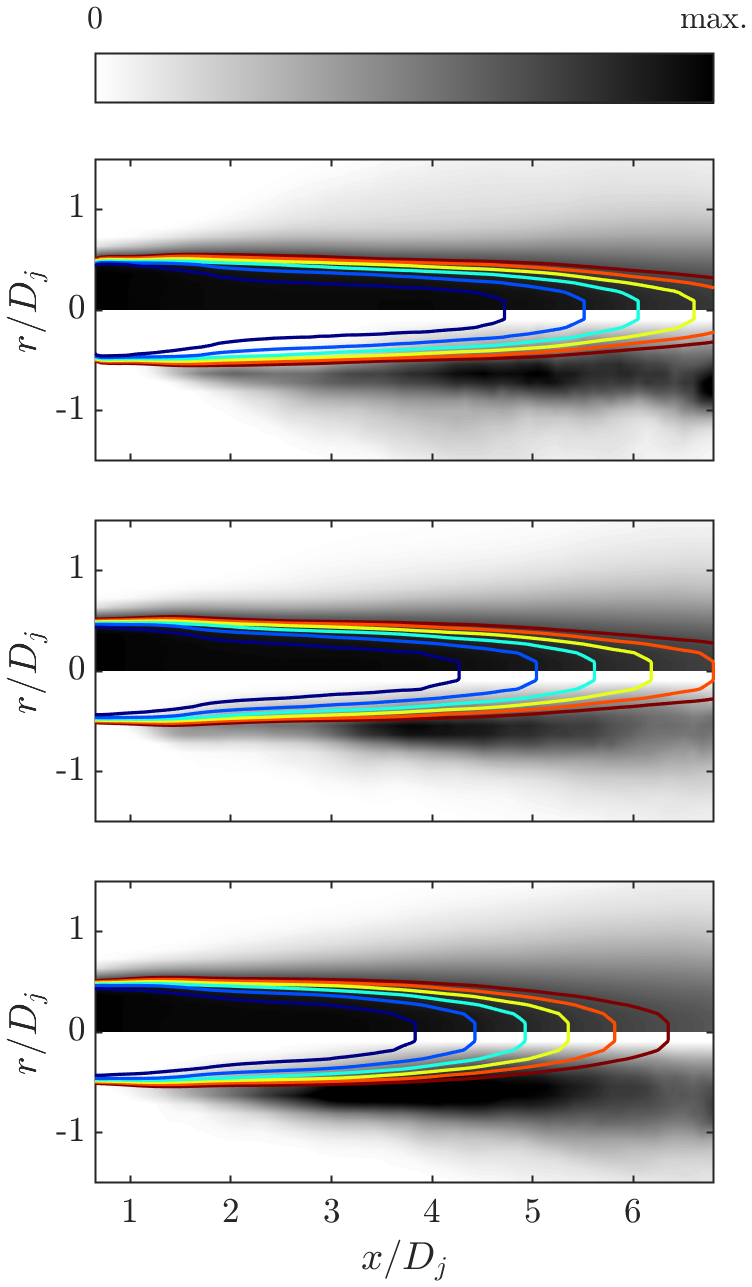}
    \caption{Time-averaged flow quantities of the unforced (top) and the forced jet with optimized axisymmetric (center) and axisym.-helical actuation (bottom). Top half shows mean axial velocity, bottom half shows Reynolds stresses. Superposed isolines indicate values of $\bar{u}_x = \left[0.95, 0.9, \dots, 0.7\right]\bar{u}_j$.}
    \label{fig:results:flow_u}
\end{figure}
% --------------------------- end fig ---------------------------------
To asses the effect of the different forcings on the time-averaged jet flow, Fig.~\ref{fig:results:flow_u} depicts corresponding fields. Each plot shows the time-averaged axial velocity field in the upper half and the axial--radial Reynolds stress component in the lower half. The superposed isolines indicate levels of time-averaged axial velocity.
All flows are statistically axisymmetric, which is a result of the construction of the control commands. In the unforced case, the axial extend of the potential core region (centerline velocity of $u_x>0.9\,u_j$) is around $6D_j$ (Fig.~\ref{fig:results:flow_u} (top)). The Reynolds stresses are evident along the shear layer with a maximum located around $x=4.5-5D_j$. When the natural axisymmetric mode is forced (Fig.~\ref{fig:results:flow_u} (center)), the distribution of Reynolds stresses is more compact and slightly shifted upstream, peaking around $x=4D_j$. The increased shear layer growth in axial direction leads to the core region being shortened to around $5D_j$. These effects are significantly more pronounced when the optimized axisym.--helical forcing is employed (Fig.~\ref{fig:results:flow_u} (bottom)). The Reynolds stress is significantly increased which leads to a shortening of the potential core to around $x=4.5D_j$. 

\begin{figure}
    \centering
    \input{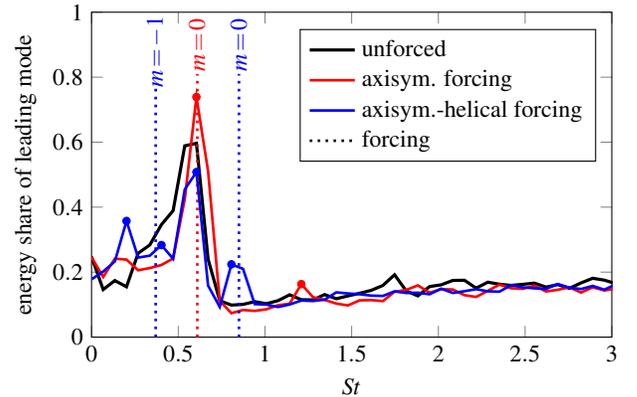}
    \caption{SPOD spectra of the unforced and forced flow. Plot shows share of leading mode w.r.t. total kinetic energy at every frequency. Dotted vertical lines indicate forcing. For highlighted Strouhal numbers Modes are shown in Figs.~\ref{fig:SPOD_modes_axial} and \ref{fig:SPOD_modes_combined}.}
    \label{fig:SPOD_spectra_forced}
\end{figure}

\begin{figure}
    \centering
    \input{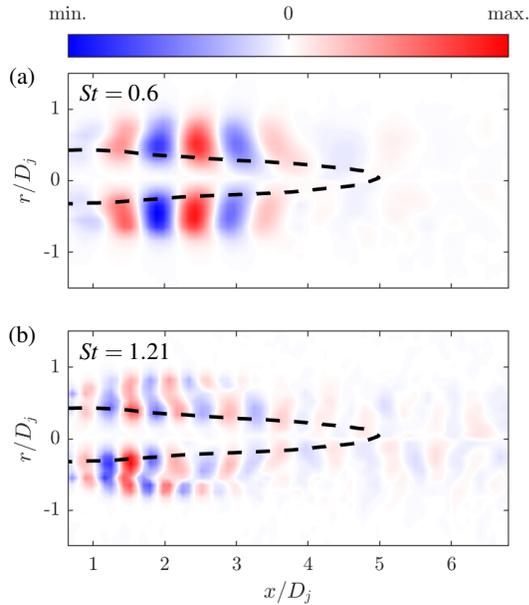}
    \caption{SPOD modes of jet flow, actuated with axisymmetric forcing. Plots show real part of radial component. Corresponding spectrum is shown in red in Fig.~\ref{fig:SPOD_spectra_forced}.}
    \label{fig:SPOD_modes_axial}
\end{figure}

% --------------------------- SPOD ---------------------------------
Using the time-resolved velocity data from PIV, we analyze the flow patterns generated from the optimal forcings by means of SPOD. The spectra of the unforced\footnote{Note that in contrast to the spectra shown in Fig.~\ref{fig:unforced_SPOD_spectrum}, the flow data is not decomposed for the unforced spectrum shown in Fig.~\ref{fig:SPOD_spectra_forced}.}, axisymmetrically forced and axisym.-helically forced flows are shown in Fig.~\ref{fig:SPOD_spectra_forced}. 
The axisymmetric actuation is optimized such that the forcing frequency is close to the natural frequency of axisymmetric structures, found in Fig.~\ref{fig:unforced_SPOD_spectrum}, a result that is consistent with other studies~\cite{crow1971jfm}. The forcing at this frequency amplifies the corresponding peak in the SPOD spectrum. Additionally, it generates a second peak at $St=1.21$, which is associated to the first harmonic of the natural mode. The associated mode shapes are shown in Fig.~\ref{fig:SPOD_modes_axial}.
The most energetic mode at $St=0.6$ resembles the naturally occurring flow pattern which seems to be particularly receptive to excitation. Careful comparison between Fig.~\ref{fig:unforced_SPOD_modes} and Fig.~\ref{fig:SPOD_modes_axial} reveals a more pronounced structure in the upstream region of the mode when the forcing is applied. The structure of the leading mode at $St=1.21$ reveals the first harmonic of the natural mode which arises as a result of non-linear effects from the fundamental one. Accordingly, the mode shape shows a symmetric vortex pattern with about half the wavelength of the fundamental. 

\begin{figure}
    \centering
    \input{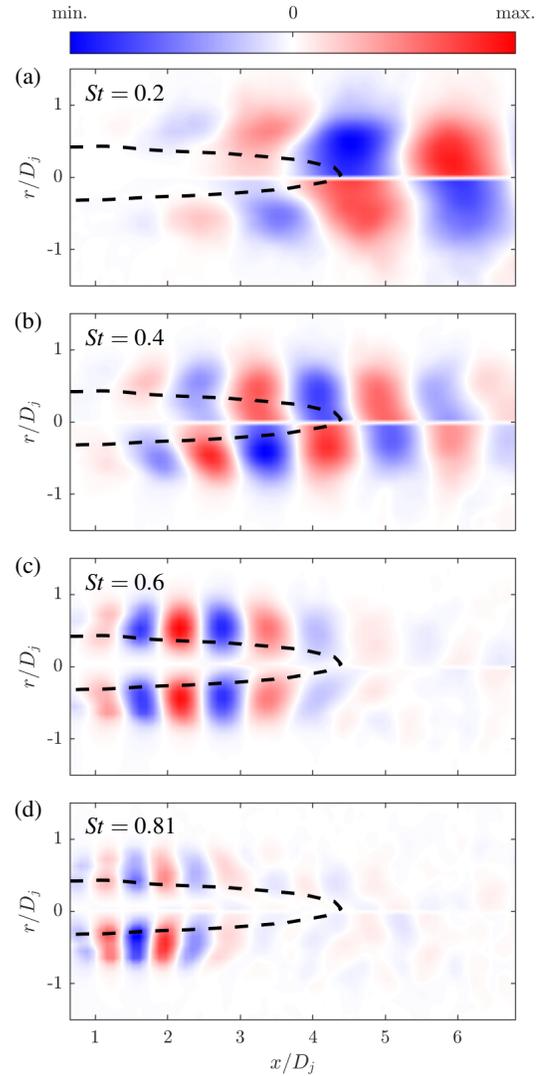}
    \caption{SPOD modes of jet flow, actuated with combined axisym.--helical forcing. Plots show real part of radial component. The corresponding spectrum is shown in blue in Fig.~\ref{fig:SPOD_spectra_forced}.}
    \label{fig:SPOD_modes_combined}
\end{figure}

The found optimal excitation with combined axisymmetric and helical mode leads to a more complex SPOD spectrum. The excitation frequencies are selected approximately symmetrically around the natural frequency of the symmetric mode ($St=0.6$) at $St_0 = 0.85$ and $St_{-1} = 0.37$. This triggers various modes in the spectrum: Around the excitation frequencies, a symmetrical and a helical flow structure oscillate (shown in Fig.~\ref{fig:SPOD_modes_combined}), corresponding to the azimuthal order of the excitation. Furthermore, the natural mode is visible at $St=0.6$. In addition, a peak in the low-frequency range is recognizable, at $St=0.2$. Since this structure was also found in the natural jet flow, we assume a high receptivity for helical modes in this frequency range. The fact that the actuation generates a peak in the spectrum with associated helical mode pattern seems to be due to a non-linear interaction between helical forcing and the natural axisymmetric modes. According to the difference in frequency and azimuthal order, a flow structure is excited in this particularly receptive range, which leads to multiple peaks in the SPOD spectrum. Thus, the optimizer combines naturally occurring modes and external excitation to excite several receptive frequency ranges with corresponding azimuthal orders. The resulting helical modes extend further in both the axial and radial directions than the axisymmetric ones (Fig.~\ref{fig:SPOD_modes_combined}), which efficiently increases shear layer growth and thus entrainment as seen in the time-averaged flow fields in Fig.~\ref{fig:results:flow_u}.

\section{Conclusion}
\label{ToC:concl}
We investigated optimal forcing patterns mode to enhance the mixing of a turbulent jet at Reynolds number $10000$ with the surrounding air. 
Eight loudspeakers distributed around the nozzle exit were used to actuate the shear layer where it is most susceptible. By controlling the phase difference between the speakers axisymmetric and helical modes could be generated which were superposed to form complex actuation commands. Optimal forcing was investigated using data-driven methods.
The parameters of the forcing were optimized using Bayesian optimization.
The optimization routine is employed for the search of optimal mode amplitudes and frequencies in three different searching spaces with increasing complexity. Additional insight into the topology of the most complex search space was generated from the data queried during optimization by means of multi-dimensional scaling. 
Two optimal forcing patterns -- axisymmetric and axisymmetric-helical and one sub-optimal forcing -- symmetric-flapping, were identified in the form of basins of attractions around local minima of the cost function. 
The optimal axisymmetric forcing is found at maximum amplitude and a Strouhal number around the characteristic frequency of the natural flow.
If an additional helical mode is included in the search, the optimal forcing combines the axisymmetric and the helical modes at a frequency ratio of $2.39$. The selected optimal frequencies make use of a non-linear interaction between natural and forced flow structures to optimize the entrainment of the surrounding air by the jet flow. The optimization in a search space of axisymmetric, helical and counter-rotating
helical forcing converges to a similar solution, turning off one of the helical modes.
The study shows how data-driven methods are suitable for optimizing and analyzing flow control. Depending on the search space, actuation patterns known from the classical literature or complex, non-trivial approaches are found through machine learning. Their interrelation can be derived from discrete data sets using topological analysis.
\section{Appendix: Investigation on optimization metrics and measurement time}

\begin{figure}
    \centering
    \includegraphics[width=0.8\linewidth]{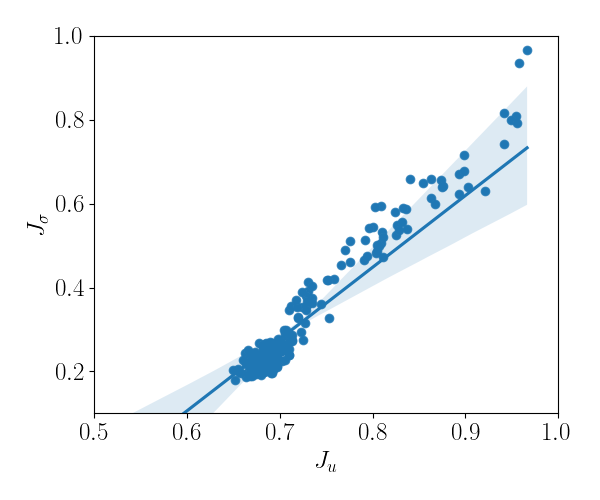}
    \caption{Comparison of centerline velocity and mean axial velocity variance as metrics for jet mixing optimization. Plot shows both metrics for 300 forcing configurations.}
    \label{fig:js}
\end{figure}

\begin{figure}
     \centering
     \includegraphics[width=0.9\linewidth]{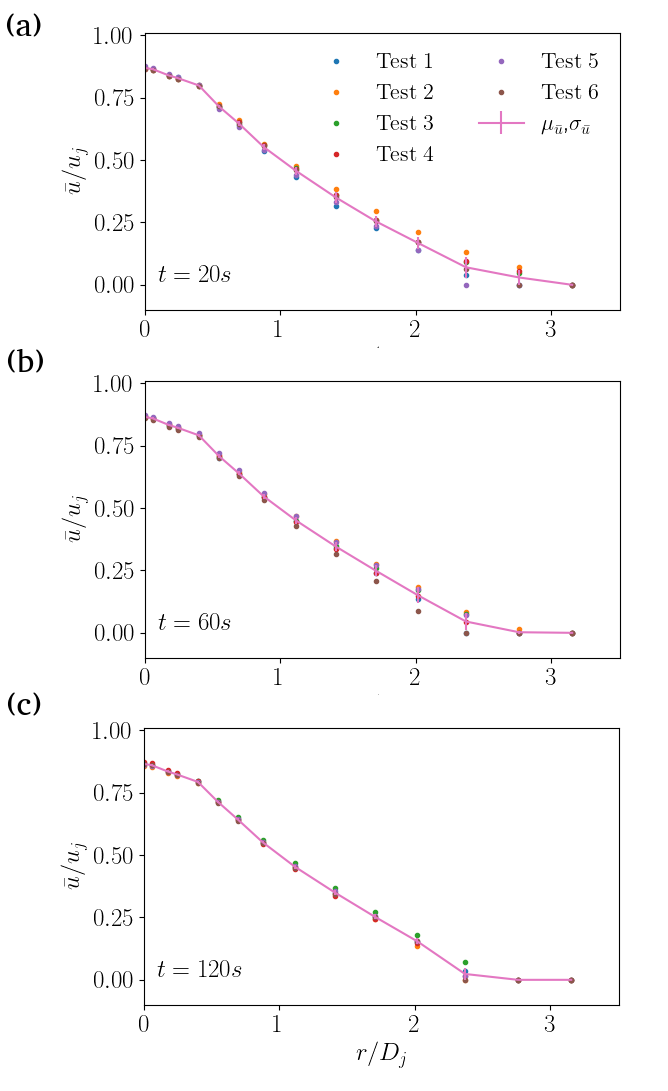}
     \caption{The measurement of streamwise velocity in the radial direction at $x=6.5D$ with different time periods.}
        \label{fig:t pitov}
\end{figure}

In addition to the centerline velocity as an objective function,
    \begin{align}
        J_{u} &= \bar{u}_{cl,x_0}.
    \end{align}
We also tested the mean axial velocity variance across the radial profile as optimization metric to represent a larger jet spread:
    \begin{align}
        J_{\sigma} &= \sum_{i=1}^{16}(\bar{u}_{x_0}^i-\bar{u}_{x_0})^2,
    \end{align}
where $\bar{u}_x^i$ is the time-averaged velocity measured from Pitot tube $i$ and $\bar{u}_x$ is the average across all $16$ tubes. We initially did so, aiming for improved robustness of the optimization by incorporating several sensor signals.
Comparing the two metrics for a set of $300$ different configurations reveals that they are indeed similarly informative, as shown in Fig. \ref{fig:js}.
However, we encountered an unfavorably low signal-to-noise ratio in the signals from the sensors in the outer shear layers, such that utilization of $J_{\sigma}$ would require significantly more measurement time per configuration.
Fig.~\ref{fig:t pitov} shows radial profiles of the streamwise velocity, acquired by averaging over time periods of different lengths, $20$, $60$, and $120$ seconds. Each measurement is repeated $6$ times to assess the measurement's variability.
% --------------------- end figure 
When the time period increases from $20$ seconds to $2$ minutes, the mean deviation across $6$ measurement trials at different locations $r/D_j$ decreases from $5\%$ to $3\%$ of the outlet velocity $u_j$.
As the figure shows, the deviation stems mainly from the tubes close to the ambient $r>2D_j$, where $\bar{u}$ is close to $0$.
In contrast, the centerline velocity $u_{cl}$ is robust enough to get a deviation lower than $1\%$, within only $20$ seconds of averaging time.
Given the above consideration,
we decided to use the centerline velocity at $x=6.5D$ as a cost function for optimization. 

\begin{acknowledgments}
    This work is supported by the National Natural Science Foundation of China under grant 12172109, 
    by the Guangdong Basic and Applied Basic Research Foundation under grant 2022A1515011492, 
    and by the Shenzhen Science and Technology Program under grant JCYJ20220531095605012. 
    We express our sincere thanks 
    to Kilian Oberleithner for sharing his expertise of the nozzle plant,
    to Antoine Blanchard and Themis Sapsis for the employed Bayesian optimization algorithm,
    % to Tianyu Wang for the employed xPDT,
    to Zhutao Jiang for discussions on experimental techniques,
    and to Artur Tyliszczak for sharing his knowledge in jet control.
\section*{Declaration of interests}
The authors report no conflict of interest.
\end{acknowledgments}

\section*{Data Availability Statement}
The data that support the findings of this study are available from the corresponding author upon reasonable request.

\section*{References}
\bibliography{literature}% Produces the bibliography via BibTeX.

\end{document}